\documentclass[journal]{IEEEtran}
\normalsize
\usepackage[pdftex]{graphicx}
  \usepackage{epstopdf}
  \DeclareGraphicsExtensions{.pdf,.jpeg,.png,.eps}
\usepackage{color}
\usepackage[cmex10]{amsmath}
\usepackage{amssymb}
\usepackage{algorithmic}
\usepackage{array}
\usepackage{mdwmath}
\usepackage{amsthm}
\usepackage{multirow}
\usepackage{mdwtab}
\usepackage{tabularx}
\usepackage{stfloats}
\usepackage{balance}
\usepackage{cite}
\makeatother

\allowdisplaybreaks
\newtheoremstyle{mystyle}
  {3pt}
  {3pt}
  {\itshape} 
  {\parindent}
  {\bfseries}
  {\upshape{:}}
  {.5em}
  {}

\usepackage{color,soul}

\theoremstyle{mystyle}

\theoremstyle{mystyle}  

\theoremstyle{mystyle}
\newtheorem{prop}{Proposition}
\usepackage[british,UKenglish]{babel}
\usepackage{subcaption}
\usepackage[font={small}]{caption}

\begin{document}

\title{Performance Analysis of SWIPT Relaying Systems in the Presence of Impulsive Noise}
\author{Lina Mohjazi,~\IEEEmembership{Member,~IEEE,} Sami Muhaidat,~\IEEEmembership{Senior Member,~IEEE,} Mehrdad Dianati,~\IEEEmembership{Senior Member,~IEEE,} Mahmoud Al-Qutayri,~\IEEEmembership{Senior Member,~IEEE,} and Naofal Al-Dhahir, ~\IEEEmembership{Fellow,~IEEE}
 \thanks{L. Mohjazi is with the Center on Cyber-Physical Systems, Khalifa University, Abu Dhabi, UAE, and also with the Preparatory Department, Khalifa University, Abu Dhabi, UAE  (e-mail: l.mohjazi@ieee.org).}
 \thanks{S. Muhaidat is with the Center on Cyber-Physical Systems, Khalifa University, Abu Dhabi, UAE, and also with the Department of Electrical and Computer Engineering, Khalifa University, Abu Dhabi, UAE (e-mail: muhaidat@ieee.org).} 
\thanks{M. Dianati is with the Warwick Manufacturing Group, University of Warwick, Coventry CV4 7AL, UK (e-mail: m.dianati@warwick.ac.uk).}
\thanks{M. Al-Qutayri is with the Department of Electrical and Computer Engineering, Khalifa University, Abu Dhabi, UAE (e-mail: mqutayri@ku.ac.ae).}
\thanks{N. Al-Dhahir is with the Electrical Engineering Department, University of Texas at Dallas, Richardson TX 75080, USA (e-mail: aldhahir@utdallas.edu).}}
\maketitle
\markboth{}{}
\begin{abstract}
We develop an analytical framework to characterize the effect of impulsive noise on the performance of relay-assisted simultaneous wireless information and power transfer (SWIPT) systems. We derive novel closed-form expressions for the pairwise error probability (PEP) considering two variants based on the availability of channel state information (CSI), namely, blind relaying and CSI-assisted relaying. We further consider two energy harvesting (EH) techniques, i.e., instantaneous EH (IEH) and average EH (AEH). Capitalizing on the derived analytical results, we present a detailed numerical investigation of the diversity order for the underlying scenarios under the impulsive noise assumption. For the case when two relays and the availability of a direct link, it is demonstrated that the considered SWIPT system with blind AEH-relaying is able to achieve an asymptotic diversity order of less than 3, which is equal to the diversity order achieved by CSI-assisted IEH-relaying. This result suggests that, by employing the blind AEH relaying, the power consumption of the network can be reduced, due to eliminating the need of CSI estimation. This can be achieved without any performance loss. Our results further show that placing the relays close to the source can significantly mitigate the detrimental effects of impulsive noise. Extensive Monte Carlo simulation results are presented to validate the accuracy of the proposed analytical framework.
\end{abstract}
\section{Introduction}
\IEEEPARstart{F}{UTURE} wireless networks are envisioned to offer an unprecedented opportunity to connect the global world via a massive number of low-power heterogeneous smart devices, enabled by the internet of Things (IoTs) \cite{Akpakwu}.  A major bottleneck for the application of such untethered nodes is their finite battery capacity, requiring the need to be recharged/replaced rather frequently. In this context, simultaneous wireless information and power transfer (SWIPT) has emerged as a promising technology to address the conflicting design goals of perpetual lifetime and uninterrupted network performance. In a SWIPT-enabled system, a wireless node is powered up by a received Radio Frequency (RF) signal and, simultaneously, information processing is carried out using the same signal \cite{Varshney2008}. 
\par SWIPT-based relaying was proposed as a promising technique to provide advantages in two fold. First, the network itself can benefit from the relays in throughput improvement, communication reliability enhancement, and coverage range extension. Second, the harvested energy can be used to charge the relay nodes, and therefore, the overall power consumption of the network may be considerably reduced \cite{Nasir2013, Mohjazi3}. From this perspective, the theoretical and implementation aspects of SWIPT relay networks have been areas of active research interest (see \cite{Rabie1,Al-habob,Fang,Ojo} and the references therein). 
\par Although there has been a growing literature on SWIPT, particularly in the context of relay networks (see e.g., \cite{Rabie1,Al-habob,Fang,Ojo} and the references therein), all research studies were based upon the classical assumption of additive white Gaussian noise (AWGN). However, many communication channels are additionally impaired by impulsive man-made electromagnetic interference or atmospheric noise encountered in various metropolitan and indoor wireless applications, such as, automotive ignition, electronic devices, household appliances, medical equipment, and industrial equipment. \cite{Blankenship, Blackard,Sanchez}. A practical foreseen scenario of such a situation is future IoTs, for instance, where nodes can be implanted in environments that are susceptible to impulsive noise such as in industrial locations or in fields close to power lines. Although these nodes are envisioned to be powered by RF energy through SWIPT to achieve advantages, such as, dual use of RF signals for information and power transfer, extended network lifetime, etc., their performance in terms of error rate is not yet studied when impulsive noise is considered. Nonetheless, it is considered as a prevalent source of performance degradation. It has been demonstrated in \mbox{\cite {Spaulding}} that communication systems designed under the AWGN assumption typically suffer from severe performance degradations when exposed to impulsive noise. This elevates the need for studying the performance of SWIPT systems, which are not only disturbed by multipath fading, but also by impulsive (non-Gaussian) noise, in order to provide pragmatic information for the system designer.
\par Several statistical models have been proposed to approximate the behaviour of impulsive noise, such as Bernoulli-Gauss \cite{Ghosh}, the symmetric alpha stable distribution \cite{Ilow}, and the Middleton's models  \cite{Middleton1, Middleton2}.  However, Middleton's models have been widely accepted to model the effects of impulse noise in communication systems due to its accuracy in approximating the behaviour of this noise over many communication channels and since its validity was confirmed by many measurement campaigns. Among the three distinct noise categories of Middleton's models, the most popular is the so-called Middleton Class-A (MCA) noise model \cite{Middleton2}. Additionally, this model presents the advantage to be a generic model which only depends on three physical parameters, namely, the noise power, the impulsive index that describes the average number of impulses during some interference time, and the Gaussian factor which resembles the ratio of the variances of the background Gaussian noise to the impulsive noise. Furthermore, the MCA noise model is characterized by a simple probability density function (PDF) expression which enables designing an optimum receiver with low complexity.
\par Several research studies in the open literature have investigated the effect of the MCA impulsive noise on conventional non-energy harvesting (EH) communication systems \cite{Alhussein1,Ping,Al-Dharrab,Gao,Schober1} and the references therein. However, these studies focus on examining the impact of impulsive noise on the process of information delivery only.  Nonetheless, SWIPT systems are characterized both by information and power delivery simultaneously.      Therefore, a thorough analysis of the effect of impulsive noise is an inevitable prerequisite for the appropriate design of impulsive noise combating mechanisms and robust receivers for such systems.
\par While most of the current literature on SWIPT systems is based upon the assumption of the classical AWGN noise assumption, there have been recent results \cite{Rabie2,Rabie3} which study the performance of a point-to-point SWIPT system under the assumption of impulsive noise following the Bernoulli-Gauss model. To the best of our knowledge, the impact of impulsive noise on the performance of SWIPT is not comprehensively understood yet, since it has not been addressed in the related open literature, which demands for a thorough investigation. We note that such an investigation is imperative for the actual realization of SWIPT and for determining the actual performance limits in terms of error rate performance.
\par Aiming to fulfil this research gap, we propose an accurate mathematical framework to analyse the pairwise error probability (PEP) performance of SWIPT relaying systems over Rayleigh fading channels subject to MCA. PEP constitutes the stepping stone for the derivation of union bounds to the error probability. It is widely used in the literature to analyse the achievable diversity order, where closed-form error probability expressions are unavailable. In particular, we assume that SWIPT relaying is enabled by a power splitting (PS) receiver architecture \cite{Nasir2013} and adopt the amplify-and-forward (AF) relaying protocol with two schemes depending on the availability of channel state information (CSI) at the relay node, namely, a CSI-assisted relaying scheme and a blind relaying scheme. Additionally, we adopt two EH techniques: EH based on average CSI (AEH) \cite{Liu} and EH based on instantaneous CSI (IEH) \cite{Mohjazi4}.  Specifically, the main contributions and results of this paper are summarized as follows:
\begin{itemize}
\item We derive novel exact closed-form PEP expressions for a two-relay dual-hop SWIPT relaying system with blind and CSI-assisted relaying schemes employing AEH and IEH.
\item The derived analytical PEP expressions are used to numerically evaluate the diversity order of the considered schemes. Specifically, we demonstrate that CSI-assisted relaying with AEH is superior to the other three relaying techniques achieving the highest diversity order of three. We further demonstrate that the lowest diversity order of two is obtained by the blind relaying scheme employing IEH suffering from cascaded fading resulting from IEH. 
\item We demonstrate that under severe noise impulsiveness, the convergence to full spatial diversity becomes slower and that the associated performance loss increases with the diversity order.
\item We demonstrate through our numerical results that for all considered relaying techniques, the best performance is achieved when the two relays are located closer to the source node than the destination node and conclude that the optimal location of the relays is independent from the noise type, i.e., MCA or AWGN.
\item Finally, a comprehensive computer-based Monte Carlo simulation study is presented to verify the accuracy of the analytical results and to further investigate several design choices within the considered relay-assisted transmission scenarios.
\end{itemize}
\par The remainder of the paper is organized as follows. In Section II, we describe the noise model and the two-relay SWIPT transmission model in conjunction with blind and CSI-assisted relaying. In Section III, we present the analytical derivations of the PEP expressions for each of the relaying techniques under consideration.  Section IV provides extensive Monte-Carlo simulation results to corroborate the analytical results and to provide detailed performance comparisons among the competing schemes for various scenarios. Concluding remarks are given in Section V. The appendices include mathematical details of the PEP derivations.  
 \par \underline{\textit{Notation}:} Bold lower case letters denote vectors. $(.)^T, (.)^*$, $\mathbb{E}[z]$, and $|z|$ stand for the transpose, conjugate, expectation of the random variable $z$, and magnitude of a complex variable $z$, respectively.  
 \begin{figure}[!t]
\centering
\includegraphics[width=3.5in]{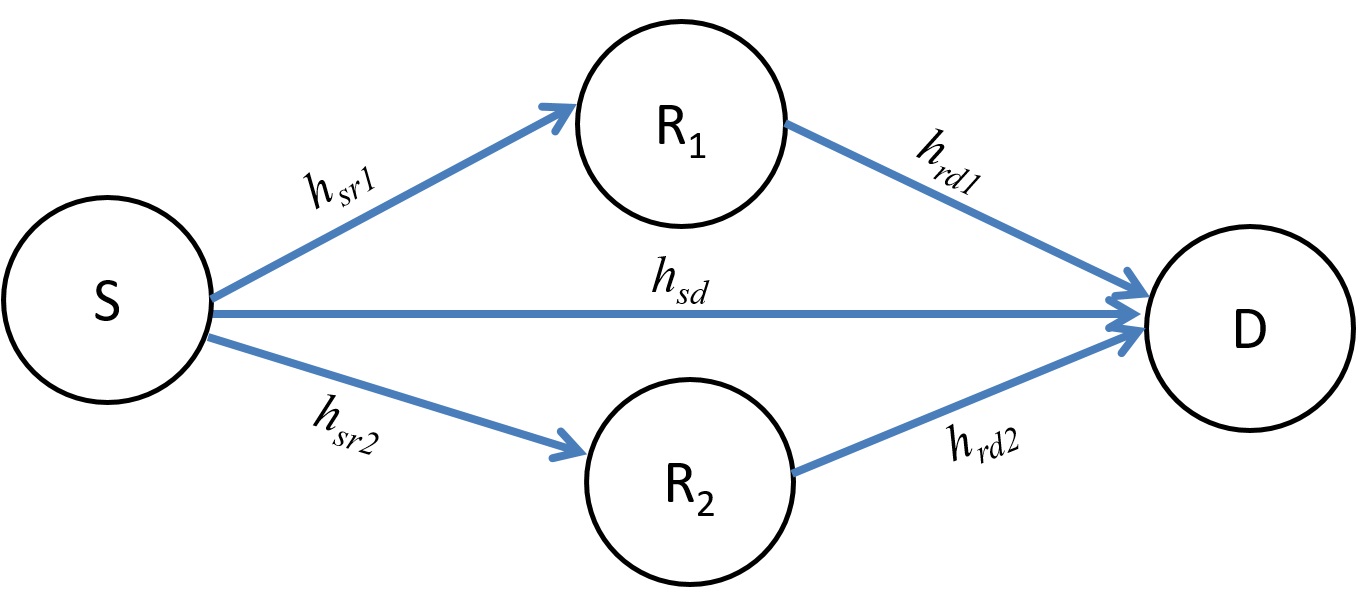}
\caption{Schematic representation of relay-assisted transmission.}
\label{blockdiagram} 
\end{figure}
\section{System Model}
\label{sec:model}
We consider a dual-hop AF SWIPT relaying system as shown in Fig.~\ref{blockdiagram}, where a source node, $S$, communicates with a destination node, $D$, via two intermediate relay nodes, $R_1$ and $R_2$. The source and the destination nodes are assumed to be energy unconstrained nodes powered by either a battery or a power grid. On the other hand, the relay nodes have no dedicated power supply and harvest energy from the received signal which is then used over the second hop. In our work, we assume that a direct link exists between the source node and the destination node. We consider the PS protocol for wireless EH, and assume that all nodes are equipped with a single antenna. We also assume that all nodes operate in the presence of impulsive noise. In what follows, we introduce the adopted noise and transmission models.
\subsection{Noise Model}
We assume that each noise sample in the $t$-th time slot at any node is given by
\begin{equation}\label{noisesample}
n(t)=n_G(t)+n_I(t),
\end{equation}where $n_G(t)$ and $n_I(t)$ denote the background zero-mean complex Gaussian noise with variance $\sigma^2_G$ and the impulsive noise with variance $\sigma^2_I$, respectively. Adopting the MCA noise model and assuming that the active interfering sources emit independently,  the PDF of the complex-valued  noise sample, given in \eqref{noisesample}, at any of the nodes can be expressed as \cite{Middleton}
\begin{equation}\label{MCApdf}
f(n(t))  =\sum_{m(t)=0}^\infty \frac{\alpha_{m(t)}}{\pi\sigma^2_{m(t)}}\text{exp}\left(-\frac{|n(t)|^2}{\sigma^2_{m(t)}}\right),
\end {equation}where\footnote{Hereafter, we drop the time index in $m(t)$ and use $m$ instead.} 
\begin{equation}\label{alpha}
\alpha_{m}=\frac{e^{-A}A^{m}}{m!},
\end{equation}with $A$ denoting the impulsive noise index that describes the average number of impulses during the interference time \cite{Middleton}. When it takes small values, i.e., $A\to 0$, it results in a highly structured and more impulsive noise, whereas it results in a near-Gaussian noise when it is large, i.e., $A\to\infty$. Furthermore, in \eqref{MCApdf}, $\sigma^2_m$ is the conditional variance given that $m$ impulses are affecting the receiver and is calculated as $\sigma_m^2=\sigma_n^2\beta_m$, where $\sigma_n^2$ denotes the mean variance of impulsive noise $n(t)$ and is equal to $N_0$ and $\beta_m$ is given by
\begin{equation}\label{beta}
\beta_m=\left(\frac{mA^{-1}+\delta}{1+\delta}\right),
\end{equation}where $\delta=\sigma^2_G/\sigma^2_I$ is called the Gaussian noise factor \cite{Middleton}, which is equal to the ratio of the variance of the background Gaussian noise component to the impulsive noise component. It is worth noting that the noise PDF in \eqref{MCApdf} reduces to the Gaussian distribution when $\delta\to\infty$ while it tends to be more impulsive when $\delta\to 0$. 
Throughout this work, we assume that $\delta>0$ which implies that the Gaussian noise component is always present. 
\par As clearly seen from \eqref{MCApdf}, the noise sample $n(t)$ in \eqref{noisesample} is not Gaussian, however, it can be viewed as conditionally Gaussian, such that, when conditioned on the state $m$,  $n(t)$ is Gaussian with zero-mean and variance $\sigma_m^2$. The sequence of states $m(t)$ is an independent and identically distributed (i.i.d) random process, and a particular state $m(t)=m$ occurs with probability $C_0=\alpha_m$, $0\leq m<\infty$, where $m(t)$ follows a Poisson distribution with parameter $A$. Therefore, it is interpreted from that the integer random variable $C_0$ is the state of the noise indicating that there is no impulse $(C_0=0)$, or impulses are present $(C_0>0)$. 
\par Although the distribution of MCA includes an infinite summation, it is completely characterized by two parameters, $A$ and $\delta$. In this work, we assume that $A$, $\delta$, and $\sigma_n^2$ are perfectly known at the receiver. In practice, these parameters can be estimated using the expectation maximization (EM) method proposed in \cite{Zabin}. We can see that the noise state probability $\beta_m$ in \eqref{beta} tends to zero as $m$ approaches infinity. Therefore, in the subsequent analysis, we truncate the sum in \eqref{MCApdf} to $M$ terms to reduce the computational complexity without compromising the performance accuracy \cite{Al-Dharrab}.
\par In this paper, we assume that the impulsive noise samples are temporally dependent during a transmission frame, following the widely used assumption in literature \cite{Delaney}. Furthermore, from the perspective of spatial dimension, we consider two models, namely, dependent and independent impulsive noise models. In Model I, which assumes spatially dependent noise samples,  the same set of interfering sources affects the destination and relay nodes together. This scenario is applicable when the destination and relay nodes are at relatively the same distance to the interfering sources \cite{Gao, Al-Dharrab}. On the contrary, in Model II, it is assumed that each of the destination and relay nodes are affected by different sets of interfering sources and therefore, their respective noise samples are spatially independent. 
\begin{figure}[!t]
\centering
\includegraphics[width=4in]{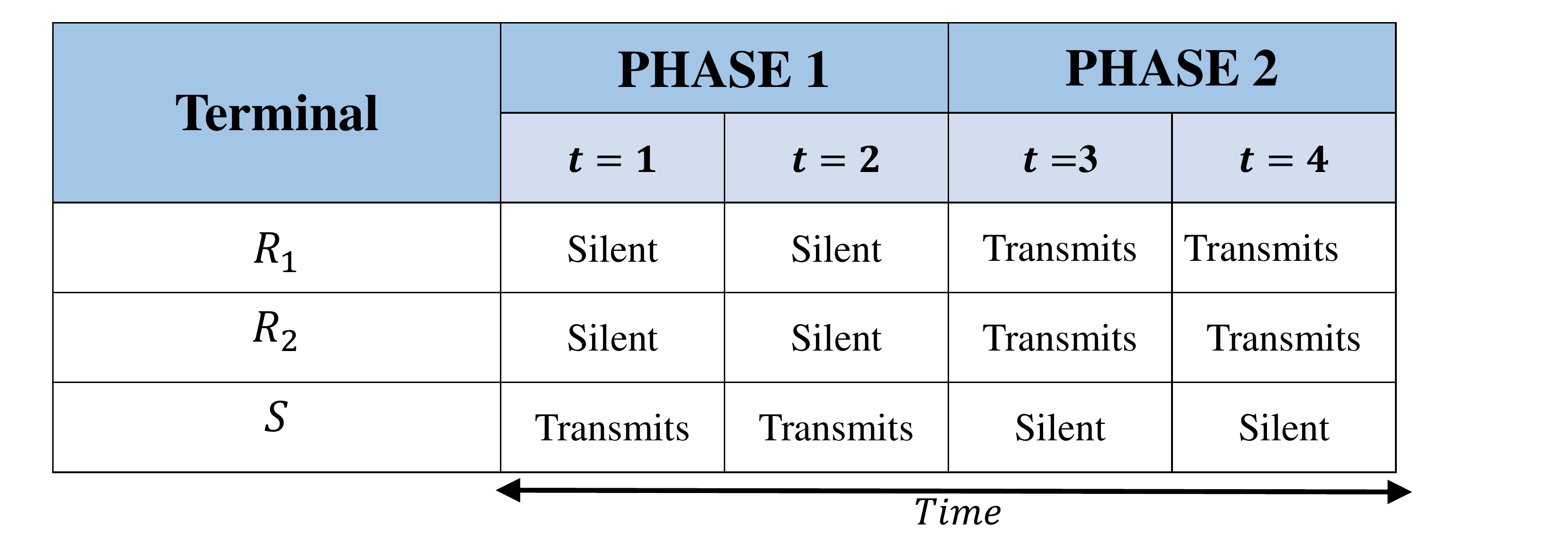}
\caption{Transmission allocation of the source node, $S$, and the two relay nodes, $R_1$ and $R_2$ over the two-Phase transmission scheme with each phase consisting of time slots.}
\label{transmissionmodelch3} 
\end{figure}
\subsection{Transmission Model}
We consider a wireless communication scenario where the source node $S$ transmits information to the destination node $D$ with the assistance of two EH relay nodes $R_1$ and $R_2$. We adopt the so-called Protocol II of \cite{Laneman, Nabar} as the relaying protocol, which is completed over two signalling intervals, namely,  \textit{Phase-1} and  \textit{Phase-2} (see Fig.~\ref{transmissionmodelch3}). We further assume that the source node communicates with the two relays and the destination nodes using the Alamouti's code \cite{Alamouti}. Specifically, the transmission of two Alamouti coded symbols is performed over four time slots ${t}=1,...,4$. During \textit{Phase-1}, spanning two time slots ${t}=1$ and ${t}=2$, the source node communicates with the relays and destination nodes. In \textit{Phase-2}, spanning two time slots ${t}=3$ and ${t}=4$, the source node remains silent, whereas the two relays employ the AF relaying technique to retransmit a scaled version of their received signals to the destination node using Alamouti coding\footnote{This protocol realizes a maximum degree of broadcasting and exhibits no receive collision \cite{Nabar}.}. Also, it is assumed that the system is perfectly synchronized at the symbol level, i.e., relays transmit at the same time \mbox{\cite{Nabar}}. Protocol II is logical in a scenario where the source node engages in data reception from another node in the network over the second time slot, thereby rendering it unable to transmit \cite{Nabar}. The implementation of the Alamouti coding scheme has been considered in the literature of SWIPT networks in \cite{Zhai,Liu2016}. We assume that the relays harvest energy from the received source signals during \textit{Phase-1}, which is then used to forward the information to the destination in \textit{Phase-2}. 
\par We further categorize the adopted AF relaying schemes based on the applied amplifying coefficient at the relay nodes, referred to as blind relaying \cite{Laneman2,Hasna} and CSI-assisted \cite{Laneman1} relaying. In the former scheme, the relays have no access to instantaneous CSI of their respective $S\to R$ links and hence, employ a fixed amplifying coefficient, which ensures that an average output power is maintained \cite{Laneman2}. While in the latter scheme, the relays use the receive CSI of their respective $S\to R$ link to ensure that the output power is limited to the power available at the relay, and therefore, a constant power is maintained for each realization \cite{Laneman1}. 
\par Let $h_{sd}$, $h_{sr,n}$ and $h_{rd,n}$, respectively denote the complex small-scale fading coefficients over the $S\to D$ link, $S\to R_n$ link from the source to the $n$-th relay, $n \in \lbrace{1,2\rbrace}$, and $R_n\to D$ link from the  $n$-th relay to the destination. These channel coefficients are modelled as i.i.d zero-mean complex Gaussian random variables (RVs) with variance 0.5 per dimension leading to the well-known Rayleigh fading channel model. It is also assumed that all channel coefficients remain constant over the block duration and vary independently and identically from one block to another. In addition to the small-scale fading, we further assume that all links are subject to large-scale path-loss that reflects the effect of the relative relays' locations on the performance of the system. Under this assumption, the received power is inversely proportional to $d_{ij}^{\lambda}$, where $d_{ij}$ is the propagation distance between transmitter $i$ and receiver $j$ and $\lambda> 2$ denotes the path-loss exponent. We set the reference distance equal to the distance from the source to the destination and assume that it is equal to unity, and hence, $d_{sr,n}= 1-d_{rd,n}, n\in\lbrace{1,2\rbrace}$. Consequently, the relative gains of $S\to R_n$ and $R_n\to D$ links are defined as $L_{sr,n}=(d_{sr,n}/d_{sd})^{\lambda}$ and $L_{rd,n}=(d_{rd,n}/d_{sd})^{\lambda}$, where $n \in \lbrace{1,2\rbrace}$. 
\par Let the two consecutive signals transmitted by the source in  \textit{Phase-1} be denoted as $s_1(t)$ and $s_2(t)$. We assume a binary phase shift keying (BPSK) signal constellation with normalized energy for the signals i.e., $\mathbb{E}[|s_p(t)|^2]=1$,  $p\in \lbrace{1,2\rbrace}$. More specifically, during the first phase, the received signals at the destination in time slots $t=1$ and $t=2$ are given by
 \begin{equation}\label{yd1}
 y_{d}(t)=\sqrt{P_s}h_{sd}s_p(t)+n_{d}(t), \quad  t=1,2,
 \end{equation}where $P_s$ is the source transmit power and $s_p(t)$, $p\in \lbrace{1,2\rbrace}$, is the symbol sent from the source in the $t$-th time interval. Also, $n_{d}(t)$ represents the overall background and impulsive noise at the destination node with conditional variance $\sigma_{m,d}^2=\beta_{m,d} N_{0_d}$, associated with the $t$-th symbol. It is recalled that the parameter $\beta_{m,d}$ depends on the occurrence of a particular random impulsive state $m$ with probability $\alpha_{m,d}$, which follows a Poisson distribution.  
\par During \textit{Phase-1}, the $n$-th relay node assigns a portion $\theta_n$ (called the PS ratio) of the received signal power in the $t$-th symbol interval for EH, and the remaining power $(1-\theta_n)$ is assigned for information processing at the information receiver.  Accordingly, the received signal at its information receiver is given by 
 \begin{equation}
\label{YIR}
 y_{r,n}(t)=\frac{\sqrt{\kappa_n P_s}}{\sqrt{L_{sr,n}}}h_{sr,n}s_p(t)+ n_{r,n}(t),
\end{equation} where $\kappa_n=(1-\theta_n)$. In this paper, we assume that $0<\theta_n<1$, corresponding to a general SWIPT system featuring both wireless information transfer and wireless EH. Furthermore, $n_{r,n}(t)$ is the overall background and impulsive noise at the $n$-th relay node associated with the $t$-th symbol, which is given by $n_{r,n}(t)=\sqrt{\kappa_n}n_{ra,n}(t)+n_{rc,n}(t)$, such that  $n_{ra,n}(t)$ and  $n_{rc,n}(t)$ are the receive antenna noise and the noise due to the RF-baseband signal conversion at the $n$-th relay, respectively, with mean variances of $N_{0_{ra,n}}$ and $N_{0_{rc,n}}$, respectively. Therefore, the conditional variance of $n_{r,n}(t)$ is $\sigma^2_{m,r,n}=\beta_{m,r,n} (\kappa_n N_{0_{ra,n}}+ N_{0_{rc,n}})$. For simplicity of the ensuing analysis, we assume that $N_{0_{ra,n}}=N_{0_{rc,n}}=N_0$.  
\par The remaining portion of the received signal at $R_n$ in the $t$-th time slot is forwarded to the energy harvester, hence, the power available at $R_n$ at the end of each of the two symbol intervals of the first phase can be expressed as 
\begin{equation}\label{Pr}
P_{r,n}=\frac{\eta_n\theta_n P_s|h_{sr,n}|^2}{L_{sr,n}},
\end{equation} with $0<\eta_n<1$ denoting the energy conversion efficiency factor at $R_n$.  It should be noted that the EH process at $R_n$ is independent of the power scaling process and it is assumed that EH is performed instantaneously. The harvested instantaneous energy is simply used as a transmit power in the second phase of transmission. Note that the assumption of instantaneous EH was adopted in \cite{Nasir2013}.
\newcounter{tempequationcounter}
\begin{figure*}[t]
\vspace*{-0.3cm}
\normalsize
 \setcounter{equation}{7}
\begin{equation}\label{yd3}
y_d(3)=\frac{\sqrt{\kappa_1 P_{r,1}P_s}}{G_{r,1}\sqrt{L_{rd,1}L_{sr,1}}}h_{sr,1}h_{rd,1} s_1(3) +\frac{\sqrt{\kappa_2 P_{r,2}P_s}}{G_{r,2}\sqrt{L_{rd,2}L_{sr,2}}} h_{sr,2}h_{rd,2} s_2(3)+\hat{n}_d(3)
\end{equation} 
\hrulefill
\vspace*{5pt}
\end{figure*} 
\newcounter{tempequationcounter1}
\begin{figure*}[t]
\vspace*{-0.3cm}
\normalsize
\begin{equation}\label{yd4}
y_d(4)=-\frac{\sqrt{\kappa_1 P_{r,1}P_s}}{G_{r,1}\sqrt{L_{rd,1}L_{sr,1}}}h_{sr,1}^*h_{rd,1} s_2(4)^* +\frac{\sqrt{\kappa_2 P_{r,2}P_s}}{G_{r,2}\sqrt{L_{rd,2}L_{sr,2}}} h_{sr,2}^*h_{rd,2} s_1(4)^*+\hat{n}_d(4),
\end{equation} 
\vspace*{4pt}
\hrulefill
\vspace*{-0.3cm}
\end{figure*}
\newcounter{tempequationcounter2}
\begin{figure*}[t]
\normalsize
\setcounter{equation}{12}
\begin{equation}\label{tildey3}
\tilde{y}_d(3)=\frac{y_d(3)}{\Omega}=\frac{\sqrt{\kappa_1 P_{r,1}P_s}}{\Omega G_{r,1}\sqrt{L_{rd,1}L_{sr,1}}}h_{sr,1}h_{rd,1} s_1(3) +\frac{\sqrt{\kappa_2 P_{r,2}P_s}}{\Omega G_{r,2}\sqrt{L_{rd,2}L_{sr,2}}} h_{sr,2}h_{rd,2} s_2(3)+\frac{\hat{n}_d(3)}{\Omega},
\end{equation}
 \hrulefill
 \setcounter{equation}{\value{equation}}
\vspace*{4pt}
\vspace*{-0.3cm}
\end{figure*}
\newcounter{tempequationcounter3}
\begin{figure*}[t]
\normalsize
\setcounter{equation}{13}
\begin{equation}\label{tildey4}
\tilde{y}_d(4)=\frac{y_d(4)}{\Omega}=-\frac{\sqrt{\kappa_1 P_{r,1}P_s}}{\Omega G_{r,1}\sqrt{L_{rd,1}L_{sr,1}}}h_{sr,1}^*h_{rd,1} s_2(4)^* +\frac{\sqrt{\kappa_2 P_{r,2}P_s}}{\Omega G_{r,2}\sqrt{L_{rd,2}L_{sr,2}}} h_{sr,2}^*h_{rd,2} s_1(4)^*+\frac{\hat{n}_d(4)}{\Omega},
\end{equation}
\hrulefill
 \setcounter{equation}{\value{equation}}
\vspace*{4pt}
\vspace*{-0.3cm}
\end{figure*}
\newcounter{tempequationcounter4}
\begin{figure*}[t]
\normalsize
\setcounter{equation}{14}
\begin{equation}\label{matrixmod}
\mathbf{y}_d=\begin{bmatrix}
y_d(1)\\y_d(2)\\y_d(3)\\y_d(4)\end{bmatrix}=\begin{bmatrix}
\sqrt{P_s}h_{sd}s_1(1)+n_{d}(1) \\
\sqrt{P_s}h_{sd}s_2(2)+n_{d}(2) \\
\frac{\sqrt{\kappa_1 P_{r,1}P_s}}{\Omega G_{r,1}\sqrt{L_{rd,1}L_{sr,1}}} h_{sr,1}h_{rd,1} s_1(3) +\frac{\sqrt{\kappa_2 P_{r,2}P_s}}{\Omega G_{r,2}\sqrt{L_{rd,2}L_{sr,2}}} h_{sr,2}h_{rd,2} s_2(3)+\tilde{n}_d(3) \\
-\frac{\sqrt{\kappa_1 P_{r,1}P_s}}{\Omega G_{r,1}\sqrt{L_{rd,1}L_{sr,1}}}h_{sr,1}^*h_{rd,1} s_2(4)^* +\frac{\sqrt{\kappa_2 P_{r,2}P_s}}{\Omega G_{r,2}\sqrt{L_{rd,2}L_{sr,2}}} h_{sr,2}^*h_{rd,2} s_1(4)^*+\tilde{n}_d(4)\end{bmatrix}.
\end{equation}
\hrulefill
 \setcounter{equation}{\value{equation}}
\vspace*{4pt}
\vspace*{-0.3cm}
\end{figure*}
\begin{figure*}[t]
\normalsize
\setcounter{equation}{18}
\begin{equation}\label{matrixmod2}
\bold{y}_d=\begin{bmatrix}
y_d(1)\\y_d(2)\\y_d(3)\\y_d(4)\end{bmatrix}=
\begin{bmatrix}
\sqrt{P_s}h_{sd}s_1(1)+n_{d}(1) \\
\sqrt{P_s}h_{sd}s_2(2)+n_{d}(2) \\
\Phi_1|h_{sr,1}| h_{sr,1}h_{rd,1} s_1(3) +\Phi_2|h_{sr,2}| h_{sr,2}h_{rd,2} s_2(3)+\tilde{n}_d(3) \\
-\Phi_1|h_{sr,1}|h_{sr,1}^*h_{rd,1} s_2(4)^* +\Phi_2|h_{sr,2}| h_{sr,2}^*h_{rd,2} s_1(4)^*+\tilde{n}_d(4)\end{bmatrix}.
\end{equation}
\hrulefill
 \setcounter{equation}{\value{equation}}
\vspace*{4pt}
\vspace*{-0.3cm}
\end{figure*}
\par During \textit{Phase-2} spanning two symbol intervals, the received signals are processed at the relay nodes using the Alamouti scheme in a distributed manner. The resulting signals are then forwarded to the destination nodes using the energy harvested in \textit{Phase-1}. Specifically, the signals received at the destination through the $R_n\to D$ links over time slots ${t}=3$ and ${t}=4$ are given by \eqref{yd3} and \eqref{yd4}, respectively, at the top of this page. In \eqref{yd3} and \eqref{yd4}, $G_{r,n}, n\in\lbrace{1,2\rbrace}$ is the scaling term at the $n$-th relay which depends on the type of amplifying coefficient deployed at $R_n$ (i.e. blind relaying or CSI-assisted relaying), which will be discussed in details in the subsequent section. This normalization does not alter the signal-to-noise ratio SNR but simplifies the ensuing presentation \cite{Nabar}. Furthermore, $\hat{n}_d(3)$ and $\hat{n}_d(4)$ are the effective noise terms associated with the third and fourth symbols, respectively, defined as
\setcounter{equation}{9}
\begin{equation}
\hat{n}_d(3)=\frac{\sqrt{P_{r,1}}}{G_{r,1}\sqrt{L_{rd,1}}}h_{rd,1}n_{r,1}+\frac{\sqrt{P_{r,2}}}{G_{r,2}\sqrt{L_{rd,2}}}h_{rd,2}n_{r,2}+n_d(3)
\end{equation} and
\setcounter{equation}{10}
\begin{equation}
\hat{n}_d(4)=\frac{-\sqrt{P_{r,1}}}{G_{r,1}\sqrt{L_{rd,1}}}h_{rd,1}n_{r,1}^{*}+\frac{\sqrt{P_{r,2}}}{G_{r,2}\sqrt{L_{rd,2}}}h_{rd,2}n_{r,2}^{*}+n_d(4).
\end{equation}\\
Assuming the so-called average power scaling (APS) \cite{Al-Dharrab}, the destination node normalizes the received signals given by \eqref{yd3} and \eqref{yd4} with
\begin{equation}
\Omega=\left(\frac{\eta_1\theta_1 P_s (\kappa_1+1) }{L_{sr,1}L_{rd,1}\mathbb{E}[|G_{r,1}|^2]}+\frac{\eta_2\theta_2 P_s (\kappa_2+1) }{L_{sr,2}L_{rd,2}\mathbb{E}[|G_{r,2}|^2]}+1\right)^{1/2},
\end{equation}resulting in \eqref{tildey3} and \eqref{tildey4}, respectively, at the top of this page. With the aforementioned signal models in mind, by letting $\tilde{n}_d(q)=\hat{n}_d(q)/\Omega, q\in\lbrace{3,4\rbrace}$, the received signal vector over four time slots is expressed as \eqref{matrixmod} at the top of the next page.
Introducing $\bold{h}=[\sqrt{P_s}h_{sd}, \sqrt{P_s}h_{sd}, D_1 h_{sr,1}^*h_{rd,1}, D_2 h_{sr,2}^*h_{rd,2}] $, where $ h_{sr,n}^*$ is chosen as $h_{sr,n}$ or  $h_{sr,n}^*$ based on the code matrix $\bold{S}$ given by
\setcounter{equation}{15}
\begin{equation}\label{codemat}
\mathbf{S}=\begin{bmatrix}
s_1(1) & 0 & 0 & 0\\
0&s_2(2)&0&0\\
0&0&s_1(3)&-s_2(4)^*\\
0&0&s_2(3)&s_1(4)^*
\end{bmatrix},
\end{equation}and $\bold{n}=[n_{d}(1), n_{d}(2), n_{d}(3), n_{d}(4)]$, the received signal vector over the whole observation period can be obtained as
\begin{equation}\label{shortsys}
\bold{y}_d=\bold{hS}+\bold{n}.
\end{equation}
\par After setting up the relay-assisted transmission model given by \eqref{matrixmod} and \eqref{shortsys}, we will now introduce the details of the signal models for blind and CSI-assisted relaying techniques.
\subsection{Blind Relaying}
Under this relaying technique, it is assumed that the $n$-th relay node does not have knowledge of its relative $S\to R_n$ link fading coefficient, therefore, it scales the received signal $y_{r,n}(t)$ by a factor of 
\begin{equation}
G_{r,n}=\sqrt{\mathbb{E}[|y_{r,n}|^2]}=\sqrt{(\kappa_n P_s/L_{sr,n})+N_{0_{r,n}})}
\end{equation} to normalize the average energy to unity \cite{Laneman2}\footnote{This power constraint is called fixed gain relaying in \cite{Hasna}}. Replacing the scaling term $G_{r,n}$, $P_{r,1}$, and $P_{r,2}$ in \eqref{matrixmod}, we can rewrite the vector form of the received signal model $\bold{y}_d$ as \eqref{matrixmod2} at the top of the next page, where $\Phi_1^2$ and $\Phi_2^2$ are given as 
\setcounter{equation}{19}
\begin{equation}\label{phi1APS}
{\Phi_{1}^2}=\frac{\eta_1\theta_1\kappa_1 P_s (P_s/N_{0_{r,1}})}{\Omega^2 L_{sr,1}^2L_{rd,1}[(\kappa_1/L_{sr,1})(P_s/N_{0_{r,1}})+1]}
\end{equation}  and 
\begin{equation}\label{phi2APS}
{\Phi_2^2}=\frac{\eta_2\theta_2\kappa_2 P_s (P_s/N_{0_{r,2}})}{\Omega^2 L_{sr,2}^2L_{rd,2}[(\kappa_2/L_{sr,2})(P_s/N_{0_{r,2}})+1]},
\end{equation} respectively.
\subsection{CSI-assisted Relaying}
Under this relaying technique, it is assumed that the relays $R_1$ and $R_2$ have knowledge about the CSI of their relative $S\to R_n$ links and accordingly, the scaling factor of the $n$-th relay becomes 
\begin{equation}
G_{r,n}=\sqrt{(\kappa_n P_s/L_{sr,n})|h_{sr,n}|^2+\beta_{I,r,n}^mN_{0_{r,n}}}.
\end{equation} Replacing this scaling term $G_{r,n}$, $P_{r,1}$, and $P_{r,2}$ in \eqref{matrixmod}, we can rewrite the vector form of the received signal at the destination as \eqref{matrixmod2}, where $\Phi_1^2$ and $\Phi_2^2$ are now written as \\
\begin{equation}\label{phi1IPS}
{\Phi_{1}^2}=\frac{\eta_1\theta_1\kappa_1 P_s (P_s/(\beta_{m,r,1} N_{0_{r,1}}))}{\Omega^2 L_{sr,1}^2L_{rd,1}[(\kappa_1/L_{sr,1})(P_s/(\beta_{m,r,1} N_{0_{r,1}}))|h_{sr,1}|^2+1]} 
\end{equation} and\\
\begin{equation}\label{phi2IPS}
{\Phi_2^2}=\frac{\eta_2\theta_2\kappa_2 P_s (P_s/(\beta_{m,r,2} N_{0_{r,1}}))}{\Omega^2 L_{sr,2}^2L_{rd,2}[(\kappa_2/L_{sr,2})(P_s/(\beta_{m,r,2}N_{0_{r,2}}))|h_{sr,2}|^2+1]},
\end{equation} \\respectively. To simplify the ensuing analysis, we assume that $N_{0}\triangleq N_{0_{sr,1}}= N_{0_{sr,2}}=N_{0_{d}}$.
\section{Pairwise Error Probability Analysis}
Based on the described noise and transmission models in the preceding section, we proceed to investigate the performance of the SWIPT relay system for each of the considered relaying techniques by deriving the PEP expressions for noise Models I and II. 

\subsection{Performance Under Noise Model I}
We start by considering the spatially dependent impulsive noise model and investigate its relative effect on the underlying SWIPT relaying system. Specifically, under Model I, the number of impulses affecting $R_1$, $R_2$, and $D$ are statistically dependent and follow the same Poisson random variable $C_0$, i.e., $\alpha_{m,d}=\alpha_{m,r,1}=\alpha_{m,r,2}=\alpha_m$. 
\par We will assume minimum distance decoding with perfect knowledge of the individual CSIs of the $S\to R_n$, $R_n\to D$, and $S\to D$ links at the receiver which is considered to be optimal when the noise is Gaussian, but is suboptimal over the impulsive noise channel \cite{Gao}. However, since the minimum distance receiver (MDR) is practical with a low detection complexity technique, we are motivated to derive its PEP performance which is mathematically tractable. 
\par Let $\bold{s}$ and $\hat{\bold{s}}$ denote the originally transmitted codeword, $\bold{s}=[s_1, s_2]$, and the erroneously-decoded codeword, $\hat{\bold{s}}=[\hat{s}_1, \hat{s}_2]$, vectors at the destination, respectively. Recalling that for the spatially dependent case, $\beta_{m,d}=\beta_{m,r,1}=\beta_{m,r,2}=\beta_{m}$, after normalising \eqref{yd3} and \eqref{yd4} by $\Omega$, then conditioned on the conditional noise variance $\beta_m$, $\tilde{n}_d(q), q\in\lbrace{3,4\rbrace},$  turns out to be a zero-mean complex Gaussian random variable with variance $\beta_m N_0$. Accordingly, the exact conditional PEP is obtained following the derivation of the conditional  PEP in the Gaussian noise case as
\begin{equation}\label{exactPEPMDR}
P(\bold{s}\to\hat{\bold{s}}|\bold{h})=\sum_{m=0}^{M-1}\alpha_m Q\left(\sqrt{\frac{d^2(\bold{s},\hat{\bold{s}})}{2\beta_m N_0}}\right),
\end{equation} where all possible realizations of the Poisson random variable $C_0$ are considered. Also, $Q(.)$ is the Gaussian-$Q$ function \cite{Rizhik} and $d^2(\bold{s},\hat{\bold{s}})$ is the Euclidean distance between $\bold{s}$, and $\hat{\bold{s}}$ written as
\begin{equation}
d^2(\bold{s},\hat{\bold{s}})=d^2_{S\to D}(\bold{s},\hat{\bold{s}})+d^2_{S\to R_1\to D}(\bold{s},\hat{\bold{s}})+d^2_{S\to R_2\to D}(\bold{s},\hat{\bold{s}})
\end{equation}
Applying the standard Chernoff bound on the $Q(.)$ function in \eqref{exactPEPMDR}, the conditional PEP can be upper bounded by \cite{Tarokh}
\begin{equation}\label{chernoff}
P(\bold{s}\to\hat{\bold{s}}|\bold{h})\leq\sum_{m=0}^{M-1}\alpha_m\text{exp}\left(\frac{-d^2(\bold{s},\hat{\bold{s}})}{4\beta_m N_0}\right).
\end{equation}
\subsubsection{PEP for Blind Relaying}
The Euclidean distance for the blind relaying scheme can be written as
\begin{align}\label{distAPS}
d^2(\bold{s},\hat{\bold{s}})=\bold{h}(\bold{S}-\hat{\bold{S}})(\bold{S}-\hat{\bold{S}})^H\bold{h}^H   \quad \quad\quad \quad \quad \quad\quad\quad \quad  \quad \quad\nonumber \\
\quad=\Delta P_s |h_{sd}|^2+\epsilon_1 \Phi_1^2|h_{sr,1}|^4|h_{rd,1}|^2+\epsilon_2\Phi_2^2 |h_{sr,2}|^4|h_{rd,2}|^2. 
\end{align} where $\Phi_1$ and $\Phi_2$ are defined in \eqref{phi1APS} and \eqref{phi2APS}, respectively, $\Delta=|s_1-\hat{s}_1|^2+|s_2-\hat{s}_2|^2$ and $\epsilon_n$ denote the eigenvalues of the codeword difference matrix $(\bold{S}-\hat{\bold{S}})(\bold{S}-\hat{\bold{S}})^H$, $n\in\lbrace{1,2\rbrace}$.  
 It is worth noting that the term $|h_{sr_n}|^4$, $n\in\lbrace{1,2\rbrace}$, appears due to the process of instantaneous EH taking place at the $n$-th relay. Henceforth, we call this relaying schemes as blind IEH-relaying. Substituting \eqref{distAPS} in \eqref{chernoff}, the PEP expression is obtained in the following proposition.
 \newcounter{tempequationcounter6}
\begin{figure*}[t]
\normalsize
\setcounter{equation}{30}
\begin{equation}\label{matrixmod3}
\bold{y}_d=\begin{bmatrix}
y_d(1)\\y_d(2)\\y_d(3)\\y_d(4)\end{bmatrix}=
\begin{bmatrix}
\sqrt{P_s}h_{sd}s_1(1)+n_{d}(1) \\
\sqrt{P_s}h_{sd}s_2(2)+n_{d}(2) \\
\Phi_1h_{sr,1}h_{rd,1} s_1(3) +\Phi_2 h_{sr,2}h_{rd,2} s_2(3)+\tilde{n}_d(3) \\
-\Phi_1h_{sr,1}^*h_{rd,1} s_2(4)^* +\Phi_2 h_{sr,2}^*h_{rd,2} s_1(4)^*+\tilde{n}_d(4)\end{bmatrix}.
\end{equation}
\hrulefill
 \setcounter{equation}{\value{equation}}
\vspace*{4pt}
\vspace*{-0.3cm}
\end{figure*}
\setcounter{equation}{28}
\begin{prop}
The unconditional PEP performance of the considered SWIPT blind IEH-relaying system in the presence of impulsive noise can be expressed in a closed-form as\\
\begin{align}\label{finalPEPmodIAPS}
P(\bold{s}\to\hat{\bold{s}})&\leq\sum_{m=0}^{M-1}\alpha_m \left(\frac{\Delta P_s}{4\beta_m N_0}+1\right)^{-1} \nonumber \\
&\times \prod_{n=1}^2\frac{1}{\sqrt{\pi}} G^{1,3}_{3,1}\left[\frac{\epsilon_n\Phi_n^2}{\beta_m N_0} \  \Big\vert \  {0.5, 0,0 \atop 0}\right],
\end{align}\end{prop}
\noindent where $ G^{m,n}_{p,q}[.\vert .]$ is the Meijer G-function defined in \cite[Eq. (8.2.1.1)]{Prudnikov}. Furthermore, $\alpha_m$ and $\beta_m$ can be calculated using \eqref{alpha} and \eqref{beta}, respectively. Note that the Meijer G-function in \eqref{finalPEPmodIAPS} can be easily and accurately computed by standard mathematical software packages such as Mathematica$^{\copyright}$, Matlab$^{\copyright}$, and Maple$^{\text{TM}}$.
\begin{IEEEproof}
See Appendix \ref{Appendix A}.
\end{IEEEproof}
\vspace*{0.1cm}
\par \textbf{Special Case (Blind AEH-relaying):}  We assume that $R_1$ and $R_2$ perform AEH which corresponds to a practical scenario where the relay nodes are equipped with a battery. Under this assumption, \eqref{Pr} which represents the power available at the $n$-th relay at the end of  \textit{Phase-1} is written as  
\begin{equation}\label{PrAEH}
P_{r,n}=\frac{\eta_n\theta_n P_s}{L_{sr,n}}.
\end{equation}Replacing \eqref{PrAEH} in  \eqref{matrixmod}, the vector form of the received signal model $\bold{y}_d$ is now given as \eqref{matrixmod3} at the top of this page. Under this scenario, $d^2(\bold{s},\hat{\bold{s}})$ is given by 
\setcounter{equation}{31}
\begin{align}\label{convdit}
d^2(\bold{s},\hat{\bold{s}})&=\Delta P_s |h_{sd}|^2 +\epsilon_1 \Phi_1^2|h_{sr,1}|^2|h_{rd,1}|^2\nonumber \\
&+\epsilon_2\Phi_2^2 |h_{sr,2}|^2|h_{rd,2}|^2. 
\end{align}It can be easily verified that \eqref{convdit} has a similar form to that in \cite[Eq. (31)]{Muhaidat1} and  \cite[Eq. (26)]{Al-Dharrab} for the conventional non-EH case. Therefore, the unconditional PEP is found as
\begin{align}\label{PEPcon}
P(\bold{s}\to\hat{\bold{s}})&\leq\sum_{m=0}^{M-1}\alpha_m\left(\frac{\Delta P_s}{4\beta_m N_0}+1\right)^{-1}\nonumber\\
&\times\prod_{n=1}^2\left(\frac{\epsilon_n\Phi_n^2}{4\beta_m N_0}\right)^{-1}  \text{exp}\left(\frac{4 \beta_m N_0}{\epsilon_n \Phi_n^2}\right)\Gamma\left(0,\frac{4 \beta_m N_0}{\epsilon_n \Phi_n^2}\right),
\end{align} where $\Gamma(a,b)=\int_b^\infty x^{a-1}\text{exp}(-x)dx$ \cite{Rizhik} denotes the upper incomplete gamma function.

\newcounter{tempequationcounter7}
\begin{figure*}[b]
\normalsize
\hrulefill
\setcounter{equation}{40}  
 \begin{align}\label{PEPmodII}
 P(\bold{s}\to\hat{\bold{s}}|\bold{h})&=\sum_{m,r,1=0}^{M-1}\sum_{m,r,2=0}^{M-1}\sum_{m,d=0}^{M-1}\left(\prod_{k=1}^3\alpha_{m,k}\right) \nonumber \\
 &\times Q\left(\frac{d^2(\bold{s},\hat{\bold{s}})}{\sqrt{2\left[\Delta P_s|h_{sd}|^2\beta_{m,d}+\epsilon_1\Phi_1|h_{sr,1}|^4|h_{rd,1}|^2\beta_{m,r,1}+\epsilon_2\Phi_2|h_{sr,2}|^4|h_{rd,2}|^2\beta_{m,r,2}\right]N_0}}\right).
  \end{align}
   \setcounter{equation}{\value{equation}}
\vspace*{4pt}
\vspace*{-0.3cm}
\end{figure*} 
 \setcounter{equation}{33} 
\subsubsection{PEP for CSI-assisted Relaying}
The Euclidean distance for the CSI-assisted relaying scheme can be written as \eqref{distAPS} where $\Phi_1^2$ and $\Phi_2^2$ are now given by \eqref{phi1IPS} and \eqref{phi2IPS}, respectively. Note that, unlike \eqref{phi1APS} and \eqref{phi2APS} for the blind relaying case, \eqref{phi1IPS} and \eqref{phi2IPS} are functions of $|h_{sr,1}|^2$ and $|h_{sr,1}|^2$, respectively. To this effect, substituting \eqref{phi1IPS} and \eqref{phi2IPS} in \eqref{distAPS}, we can write $d^2(\bold{s},\hat{\bold{s}})$ as
\begin{align}\label{distIPS}
d^2(\bold{s},\hat{\bold{s}})&= \Delta P_s |h_{sd}|^2 +\epsilon_1 \zeta_1 \frac{|h_{sr,1}|^4|h_{rd,1}|^2}{\xi_1 |h_{sr,1}|^2+1}\nonumber \\
&+\epsilon_2\zeta_2 \frac{|h_{sr,2}|^4|h_{rd,2}|^2}{\xi_2 |h_{sr,2}|^2+1}   \nonumber \\
& =\Delta P_s |h_{sd}|^2 +\epsilon_1 \zeta_1 \frac{X_1^2Y_1}{\xi_1 X_1+1}+\epsilon_2\zeta_2 \frac{X_2^2 Y_2}{\xi_2 X_2+1}. 
\end{align} where $\xi_n=[(\kappa_n/L_{sr,n})(P_s/(\beta_I^m N_0))]$, $\zeta_n$ is given as \\
\begin{equation}\label{zeta}
\zeta_n =\frac{\eta_n\theta_n\kappa_n P_s (P_s/(\beta_m N_0))}{\Omega^2 L_{sr,n}^2L_{rd,n}}, \quad n\in \lbrace{1,2\rbrace}.
\end{equation}and $X_n\triangleq|h_{sr,n}|^2$,  $Y_n\triangleq|h_{rd,n}|^2$. To obtain an expression for the PEP for the CSI-assisted IEH-relaying, let $Z_n=U_n/V_n$, where $U_n=X_n^2Y_n$ and $V_n=\xi_n X_n+1, n\in \lbrace{1,2\rbrace}$. Then, one could obtain the unconditional PEP by taking the expectation of \eqref{chernoff} with respect to the RVs $|h_{sd}|^2, Z_1$ and $Z_2$. In the following proposition, we derive the unconditional PEP expression. \\
\begin{prop}
The unconditional PEP performance of SWIPT CSI-assisted IEH-relaying system in the presence of impulsive noise can be expressed as 
\vspace*{-0.5cm}
\end{prop}
\begin{align}\label{pepIEHIPS}
P(\bold{s}\to\hat{\bold{s}})&\leq\sum_{m=0}^{M-1}\alpha_m\left(\frac{\Delta P_s}{4\beta_m N_0}+1\right)^{-1}\nonumber \\
&\prod_{n=1}^2\frac{1}{2B_n\psi_n}\left[\text{exp}(\Lambda_n)\text{Ei}(\Lambda_n)\text{D}_1+\text{exp}(\Psi_n)\text{Ei}(\Psi_n)\text{D}_2\right], 
\end{align}where $\text{Ei}(.)$ is the exponential integral function \cite{Prudnikov}, $\psi=\sqrt{\xi_n^2-4B_n}$ where $\xi_n$ is defined before \eqref{zeta}, $\text{D}_1=-\xi_n^2-\xi_n\psi_n+2 B_n$, $\text{D}_2=\xi_n^2-\xi_n\psi_m-2 B_n$, and $\Lambda_n$  and $\psi_n$ are given by 
\begin{equation}
\Lambda_n=\frac{\xi_n+\psi_n}{2B_n},
\end{equation}and
\begin{equation}
\Psi_n=\frac{\xi_n-\psi_n}{2B_n},
\end{equation}respectively.
\begin{IEEEproof}
See Appendix \ref{Appendix B}.
\end{IEEEproof}\vspace*{0.1cm}
\par \noindent \textbf{Special Case (Asymptotic PEP in high SNR):} To give more insight into the PEP performance, we consider the high SNR assumption, i.e., $\xi_n\to \infty$. Under this assumption, the second factor in the denominators of \eqref{distIPS} can be negligible. Consequently, $d^2(\bold{s},\hat{\bold{s}})$ in \eqref{distIPS} is reduced to \eqref{convdit}, yielding the PEP expression to be given as \eqref{PEPcon}. \\
\par \noindent \textbf{Special Case (CSI-assisted AEH-relaying):} Similar to the blind-relaying scenario, we assume here that $R_1$ and $R_2$ perform average EH. Under this assumption, the power available at the $n$-th relay at the end of  \textit{Phase-1} is given by \eqref{PrAEH}. Hence, we get
\begin{align}\label{distIPS2}
d^2(\bold{s},\hat{\bold{s}})&= \Delta P_s |h_{sd}|^2  +\epsilon_1 \zeta_1 \frac{|h_{sr,1}|^2|h_{rd,1}|^2}{\xi_1 |h_{sr,1}|^2+1}\nonumber \\
&+\epsilon_2\zeta_2 \frac{|h_{sr,2}|^2|h_{rd,2}|^2}{\xi_2 |h_{sr,2}|^2+1}, 
 \end{align}where $\xi_n$ and $\zeta_n$ are given below \eqref{distIPS}. Substituting \eqref{distIPS2} in \eqref{chernoff}, followed  by taking the expectation with respect to $|h_{sd}|^2,  |h_{sr,1}|^2,  |h_{sr,2}|^2,  |h_{rd,1}|^2$ and $|h_{rd,2}|^2$, the unconditional PEP is given in the following proposition.
 \begin{prop}
The unconditional PEP performance of SWIPT CSI-assisted AEH-relaying system can be expressed as
\end{prop}
 \vspace*{-0.5cm}
\begin{align}\label{PEPAEHIPS}
P(\bold{s}\to\hat{\bold{s}})\leq\sum_{m=0}^{M-1}\alpha_m\left(\frac{\Delta P_s}{4\beta_m N_0}+1\right)^{-1}\quad\quad\quad\quad\quad\quad\quad\quad\quad\quad\quad\quad\nonumber  \\
\quad\quad\times \prod_{n=1}^2\left(\gamma_n^{-1}G^{1,2}_{2,1}\left[\gamma_n \  \big\vert \  {1, 1 \atop 1}\right]+\xi_n \gamma_n^{-2}G^{1,2}_{2,1}\left[\gamma_n \  \big\vert \  {1, 2 \atop 2}\right]\right). \quad\quad\quad\quad
\end{align}
\begin{IEEEproof}
See Appendix \ref{AppendixC}.
\end{IEEEproof}
 \vspace*{0.3cm}
\noindent It is noted from each of \eqref{finalPEPmodIAPS},  \eqref{PEPcon},  \eqref{pepIEHIPS}, and  \eqref{PEPAEHIPS} that these expressions include the conventional AWGN assumption as a special case. It is recalled from \eqref{beta} that as $\delta\to\infty$, $\beta_m$ converges to 1. Therefore, the summation in \eqref{finalPEPmodIAPS},  \eqref{PEPcon},  \eqref{pepIEHIPS}, and  \eqref{PEPAEHIPS} will be equal to 1, reducing these expressions to the PEP expressions for the conventional AWGN case. It is worth mentioning that due to the presence of the summation term, in the above mentioned expressions, the convergence to asymptotic diversity order under impulsive noise is slower compared to the AWGN case. 
\par Based on the previously derived PEP expressions the diversity order \mbox{$D$} can be computed as \cite{Simon}
\begin{equation}\label{dorder}
D= -\lim_{\text{SNR} \rightarrow \infty}\frac{\text{log}\left(P(\bold{s}\to\hat{\bold{s}})\right)}{\text{log}\left(\text{SNR}\right)}.
\end{equation}Since the only source of power is, \mbox{$P_s$}, the performance of the entire system is parametrized by SNR \mbox{$\triangleq P_s/N_0$}. Using \mbox{\eqref{dorder}},  the diversity order of blind IEH-relaying, blind AEH-relaying, CSI-assisted IEH-relaying, and CSI-assisted AEH-relaying are numerically evaluated by substituting \mbox{\eqref{finalPEPmodIAPS}, \eqref{PEPcon}, \eqref{pepIEHIPS}, and \eqref{PEPAEHIPS} in \eqref{dorder}}, respectively.
\subsection{Performance Under Noise Model II}
In the following, we will study the performance of the considered SWIPT relay system under the assumption of spatially independent impulsive noise model, where $R_1$, $R_2$, and $D$ nodes are affected by statistically independent number of impulses, respectively, following Poisson random variables $C_{r,1}$, $C_{r,2}$, and $C_d$, i.e.,  $\alpha_{m,d}, \alpha_{m,r,1}$, and $\alpha_{m,r,2}$ may not necessarily be equal. In particular, the conditional variances  $\beta_{m,d}$, $\beta_{m,r,1}$, and $\beta_{m,r,2}$ are not necessarily equal. To address the independence in the number of impulses occurring at $R_1$, $R_2$, and $D$, the PEP expression has to be averaged over all possible realizations of each of $C_{r,1}$, $C_{r,2}$, and $C_d$, and thus, the conditional PEP is given by \eqref{PEPmodII} at the bottom of this page, where $\alpha_{m,1}=\alpha_{m,d}$, $\alpha_{m,2}=\alpha_{m,r,1}$, and $\alpha_{m,3}=\alpha_{m,r,2}$. 
\par To evaluate the unconditional exact PEP for each of the relaying schemes described in Section II, the expression in \eqref{PEPmodII} has to be averaged over fading coefficients $\bold{h}$, which is mathematically intractable. However, we can obtain an approximate expression for the conditional PEP in \eqref{PEPmodII} by setting $\beta_{m,r,1}=\beta_{m,r,2}=\beta_{m,d}=\bar{\varphi}$, which denotes the average number of impulses affecting $R_1$, $R_2$, $D$ nodes during a transmission frame and is given by \cite{Al-Dharrab}
 \setcounter{equation}{41} 
  \begin{equation}
  \bar{\varphi}=\frac{2(\beta_{m,r,1}+\beta_{m,r,2})+4\beta_{m,d}}{8}. 
  \end{equation}Then, by using the Chernoff upper bound, taking the expectation over the fading coefficients $\bold{h}$, and following the same line of analysis performed in the derivation of the PEP expressions of the blind and CSI-assisted relaying schemes under noise Model I, the PEP performance under noise Model II can be evaluated.
\section{Numerical and Simulation Results}\label{sec:results}
In this section, we provide a variety of numerical and Monte Carlo simulation results to validate the accuracy of the proposed analytical framework and to compare the performance of the considered blind and CSI-assisted relaying techniques employed for a SWIPT relaying system under the MCA noise Models I and II. The term Monte Carlo simulations refers to the use of actual fading channel variates with a number of repetitions of $10^6$ trials. We further assume that the two relays are  located on the straight line between the source and the destination nodes. Unless otherwise specified, in order to study various degrees of noise impulsiveness, we use three sets of values for the impulsive noise parameters $A$ and $\delta$: $(A,\delta)=(1,0.1)$,  $(A,\delta)=(0.1,0.1)$, and $(A,\delta)=(0.001,0.1)$ to represent \textit{near-Gaussian} (NG),  \textit{moderately impulsive} (MI), and  \textit{highly impulsive} (HI) noise channels, respectively, which fit well within the practical ranges of $A$ and $\delta$  \cite{Delaney}.
\par Unless otherwise stated, we set the EH efficiency factor $\eta_1=\eta_2=0.3$ as a worst case, capturing the effects of low-cost hardware, the PS factors $\theta_1=\theta_2=0.5$, the normalized distances of both relays for their respective $S\to R$ links are set to $d_{sr,1}=d_{sr,2}=0.5$, the source transmission power $P_s=1$ Watt and the path-loss exponent $\lambda=2.7$ \cite{Nasir2013}. The simulation parameters are summarized in Table \ref{t2}.
\begin{table}[!t]
\centering
\caption{Simulation Parameters}
\label{t2}
\begin{tabular}{|l|c|c|}
\hline
\multicolumn{1}{|c|}{\textbf{Name}} & \textbf{Symbol} & \textbf{Value}               \\ \hline
Impulsive noise index               & \textit{A}      & 1 (NG), 0.1 (MI), 0.001 (HI) \\ \hline
Gaussian noise factor               & $\delta$        & 0.1                          \\ \hline
Number of interfering sources       & \textit{M}      & 5                            \\ \hline
PS ratio for $R_1$                  & $\theta_1$      & 0.5                          \\ \hline
PS ratio for $R_2$                  & $\theta_2$      & 0.5                          \\ \hline
Normalized $S\to R_1$ distance      & $d_{sr,1}$      & 0.5                          \\ \hline
Normalized $S\to R_2$ distance      & $d_{sr,2}$      & 0.5                          \\ \hline
Path-loss exponent                  & $\lambda$       & 2.7                          \\ \hline
EH efficiency of $R_1$              & $\eta_1$        & 0.3                          \\ \hline
EH efficiency of $R_2$              & $\eta_2$        & 0.3                          \\ \hline
Source transmit power               & $P_s$           & 1 Watt                       \\ \hline
\end{tabular}
\end{table}

  \begin{figure}[!t]
\centering
   \includegraphics[width=3.7in]{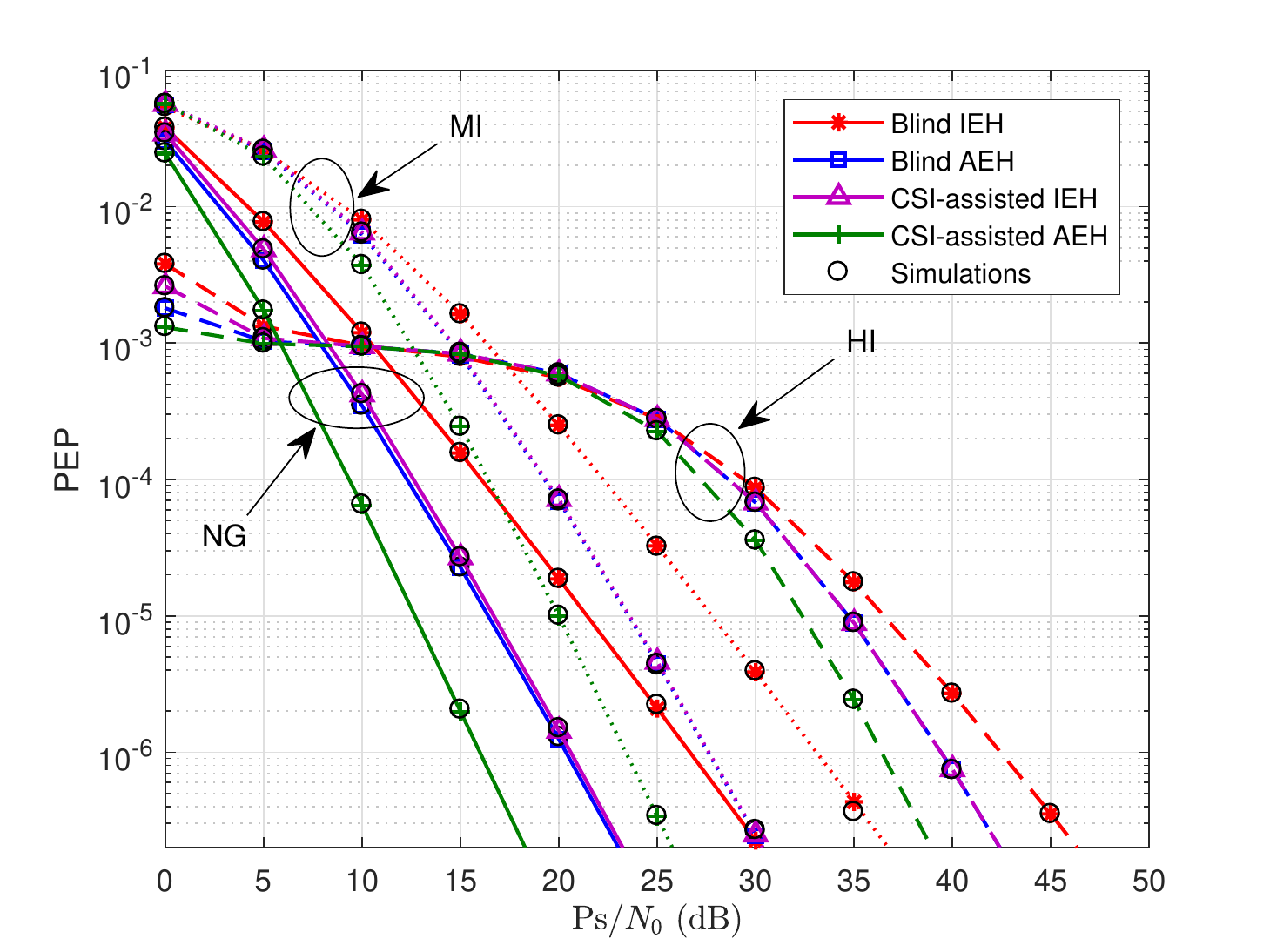}
  \caption{PEP performance with respect to SNR for blind and CSI-assisted relaying techniques over Rayleigh fading channels in the presence of HI, MI, and NG MCA noise for Model I.}
   \label{f3} 
 \vspace*{-0.5cm}
\end{figure} 
\par In Fig. \ref{f3}, we compare the PEP performance of the blind and CSI-assisted relaying techniques when IEH or AEH are considered under three MCA noise environments, namely, HI, MI, and NG, for noise Model I. Furthermore, to evaluate the accuracy of our mathematical models presented in \eqref{finalPEPmodIAPS}, \eqref{PEPcon},  \eqref{pepIEHIPS}, and \eqref{PEPAEHIPS}, we present in Fig. \ref{f3} the corresponding Monte Carlo simulation results. It is observed that the analytical PEP curves are in full agreement with the simulation results over the whole SNR operating range. This finding directly reflects the accuracy of our proposed mathematical framework and its effectiveness in quantifying the performance of the considered relaying techniques under MCA noise. It is illustrated in Fig. \ref{f3} that for all the studied relaying techniques, namely, blind relaying with IEH or AEH and CSI-assisted relaying with IEH or AEH, the PEP curves undergo a flattening when the SNR is between 5 - 20 dB under the HI noise environment, which dramatically differs from those of the NG noise environment. This behaviour is also reported for non-EH systems cooperative systems \cite{Al-Dharrab, Alhussein1} and is due to the fact that the tails of the PDF of the MCA noise becomes wider as the impulsive index $A$ decreases. However, as $A$ increases, the tails of the MCA density asymptotically approach those of a Gaussian density, resulting in the behaviour observed for the PEP performance. Moreover, for the three noise scenarios, it is shown that the performance exhibited by the CSI-assisted AEH-relaying is superior to that of the other three relaying techniques. Although, CSI-assisted relaying schemes are intuitively expected to outperform their blind relaying counterparts, our results show that the CSI-assisted IEH-relaying and blind AEH-relaying technique experience identical PEP performance. This indicates that the extra power consumption, resulting from CSI estimation, can be avoided without causing performance loss. However, this comes at the expense of requiring a battery to perform AEH.
\par
\begin{figure}[!t]
\centering
   \includegraphics[width=3.2in]{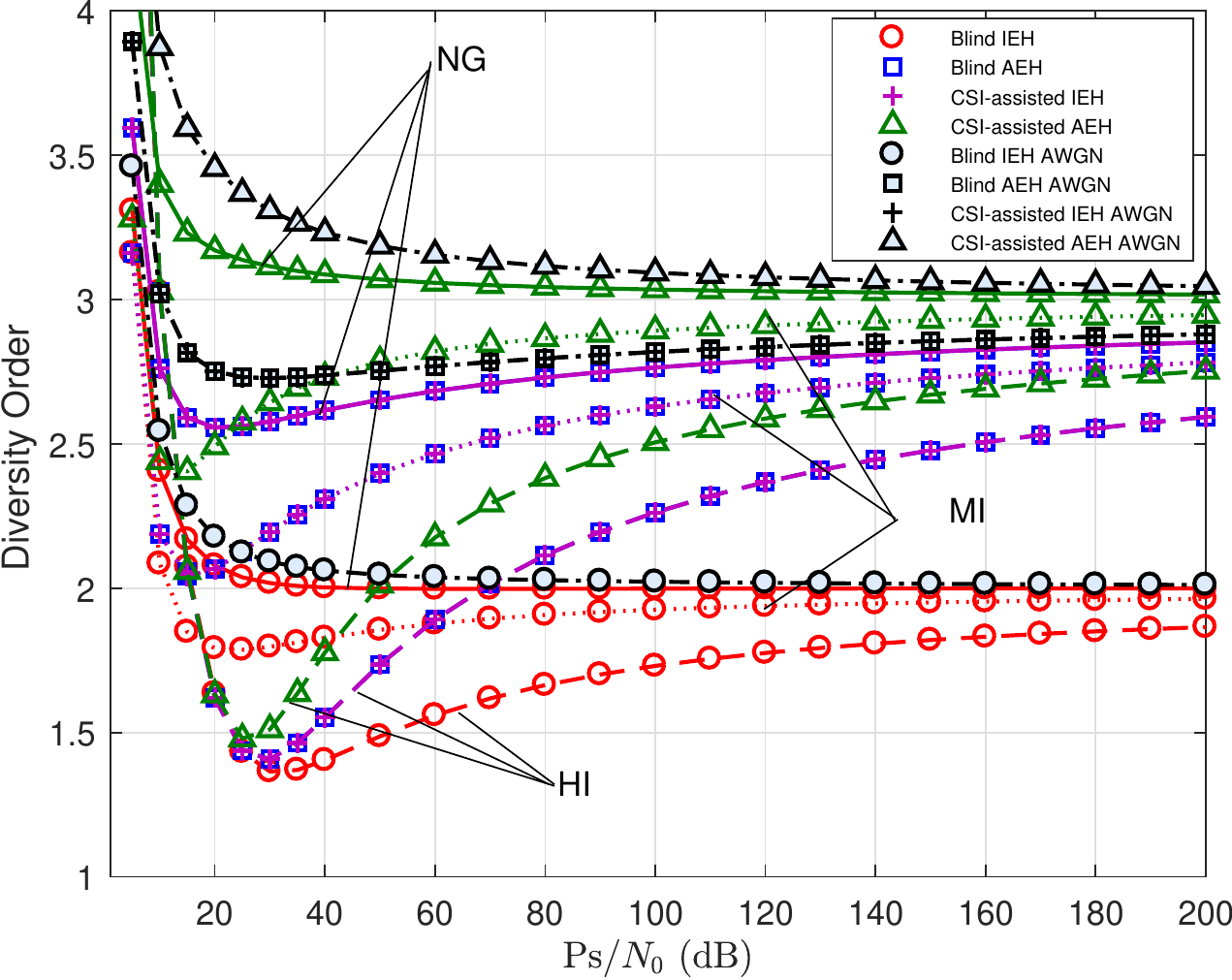}
  \caption{Diversity order of blind and CSI-assisted relaying schemes in the presence of HI, MI, and NG noise for Model I.}
   \label{f4} 
\end{figure}

 In an attempt to gain more insights about the performance of the considered relaying techniques, we investigate the achievable diversity order. Specifically, in Fig. \ref{f4}, we utilize the expressions obtained in \eqref{finalPEPmodIAPS}, \eqref{PEPcon},  \eqref{pepIEHIPS}, and \eqref{PEPAEHIPS} to calculate the diversity order, defined as the negative of the asymptotic slope of the PEP on a log-log scale \cite{Tarokh}. The achievable diversity order in the presence of the well-known AWGN case is included as a benchmark.    
\par Fig. \ref{f4} demonstrates that the CSI-assisted AEH-relaying scheme enables the system to achieve the highest diversity order $(d=3$, at NG), whereas the lowest $(d=2$, at NG) is obtained by the blind IEH-relaying scheme, where the performance is severely degraded. This is due to the effect of cascaded fading resulting from IEH. Meanwhile, the attainable diversity order for both the blind AEH-relaying and CSI-assisted IEH-relaying is identical $(d=2.85$, at NG). In Table \ref{t1}, we present the achievable diversity order levels observed by the investigated  four relaying techniques under the three MCA noise environments, along with the corresponding AWGN case. It is noted that for all the studied relaying techniques: as the impulsive noise index $A$ becomes smaller, (i.e., the noise becomes highly impulsive), the convergence to full spatial diversity, represented by the AWGN case, becomes slower. This can be attributed to the performance loss introduced by the impulsive nature of the noise incurred by the MDR.  Additionally, the full diversity order of all relaying techniques in the MCA noise environments are not realized due to the noise impulsiveness severity. Interestingly, as the noise impulsiveness level increases from NG to HI, the associated performance loss increases with the diversity order. This result is consistent with the conclusion reported in \cite{Gao} for a non-cooperative non-EH wireless communication system. 
\begin{table}[h]
\centering
\caption{Achievable diversity order under MCA noise and AWGN}
\label{t1}
\begin{tabular}{|l|l|l|l|l|}
\hline
\multicolumn{1}{|c|}{\multirow{2}{*}{Relaying Technique}} & \multicolumn{3}{c|}{MCA Noise} & \multicolumn{1}{c|}{\multirow{2}{*}{AWGN}} \\ \cline{2-4}
\multicolumn{1}{|c|}{}                                    & HI       & MI       & NG       & \multicolumn{1}{c|}{}                      \\ \hline
Blind IEH                                                 & 1.86        & 1.96       &1.99        &\multicolumn{1}{c|}{2}     

                                        \\ \hline
Blind AEH                                                 &2.59          &2.78          &2.85          & \multicolumn{1}{c|}{2.87}                                           \\ \hline
CSI-assisted IEH                                          &2.59          &2.78          &2.85          &\multicolumn{1}{c|}{2.87}                                            \\ \hline
CSI-assisted AEH                                          &2.38          &2.86          &3          &  \multicolumn{1}{c|}{3}                                          \\ \hline
\end{tabular}
\end{table}
\par
\begin{figure}[!t]
\centering
   \includegraphics[width=3.5in]{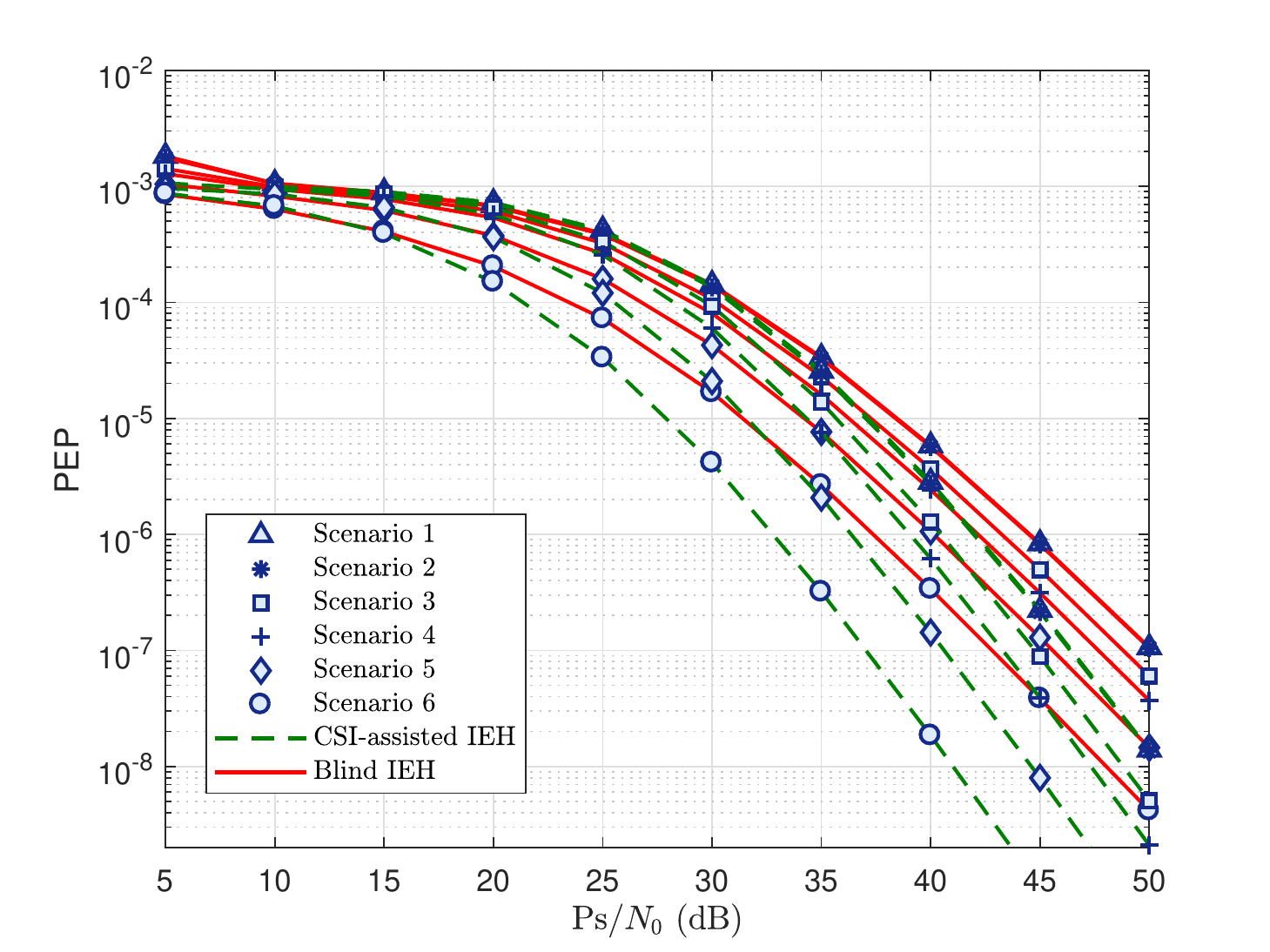}
  \caption{PEP performance with respect to SNR for various relay locations over HI noise under Model I.}
   \label{f5} 
\vspace*{-0.6cm}
\end{figure}

  To explore the effect of the relays' locations on the PEP performance of the considered blind and CSI-assisted relaying techniques with IEH,  we illustrate in Fig. \ref{f5} the performance of the Alamouti-based scheme, under the assumption of HI MCA noise.  This study is conducted for six distinct scenarios of the geometrical layout of the two relays:
\begin{itemize}
\item Scenario 1: $d_{sr,1}$=0.8 and $d_{sr,2}$=0.8, 
\item Scenario 2: $d_{sr,1}$=0.5 and $d_{sr,2}$=0.8,
\item Scenario 3: $d_{sr,1}$=0.2 and $d_{sr,2}$=0.8,
\item Scenario 4: $d_{sr,1}$=0.5 and $d_{sr,2}$=0.5,
\item Scenario 5: $d_{sr,1}$=0.5 and $d_{sr,2}$=0.2,
\item Scenario 6: $d_{sr,1}$=0.2 and $d_{sr,2}$=0.2.
\end{itemize}
It is shown in Fig. \ref{f5} that the best performance for both blind and CSI-assisted IEH-relaying schemes is exhibited by Scenario 6, where both relays are close to the source, while locating the two relays close to the destination represented by Scenario 1 leads to the worst performance. This is expected, since the power available at the relay nodes resulting from EH during \textit{Phase-1}, as defined in \eqref{Pr}, is inversely proportional to the distance between the source and the relay node. Specifically, as $d_{sr,n}, n\in\lbrace{1,2\rbrace}$ increases, both the harvested energy and the received signal strength at the relay node decrease due to the increased path-loss, and consequently, deteriorating the performance. A similar result is noted for both of the relaying techniques when AEH is employed, however their performance is not plotted to avoid repetition. 
\par This observation suggests the support for the conclusion in \cite{Nasir2013} for a SWIPT wireless cooperative systems under the general AWGN noise assumption. On the contrary, this finding is different from the conventional case where EH is not considered at the relays \cite{Al-Dharrab}, wherein the best performance is attained by Scenario 4, where both relays are equidistant from the source and destination nodes and the worst performance is observed in Scenario 3 where one of the relays is placed closer to the source node and the other is placed closer to the destination node. The aforementioned result along with the ones reported in \cite{Nasir2013} and \cite{Al-Dharrab} lead us to conclude that the optimal position of the relays in a SWIPT relaying system may be independent from the channel noise type. 
\par
\begin{figure}[!t]
\centering
   \includegraphics[width=3.7in]{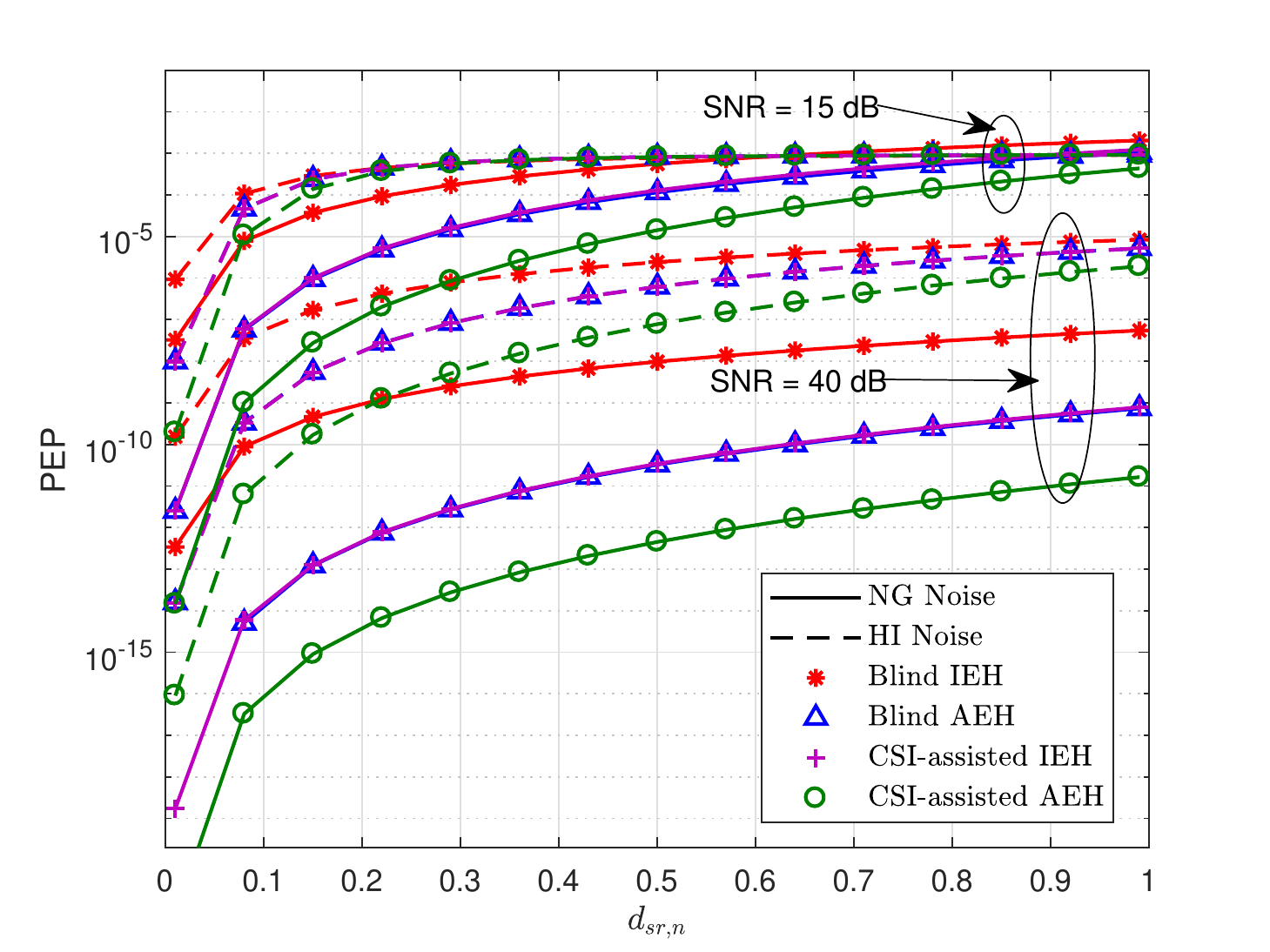}
  \caption{PEP performance with respect to the normalized distances $d_{sr,n}, n\in \lbrace{1,2\rbrace}$  over NG and HI noise under Model I.}
   \label{f6} 
\vspace*{-0.3cm}
\end{figure}

 Remarkably,  for both blind and CSI-assisted relaying techniques, as the two relays become closer to the source the flat region observed in the case of HI noise is significantly diminished, thereby, considerably outperforming the non-EH case presented in \cite{Al-Dharrab} from this perspective. Therefore, the results obtained in this examination are two fold. First, it is noted that EH relaying systems are more robust towards impulsive noise. Second, the location of the relays plays a crucial role in the underlying system performance. Further examinations of the impact of the relays' location on the system performance are carried out in Fig. \ref{f6}. 
\par Fig. \ref{f6} depicts the PEP performance of blind and CSI-assisted relaying for both IEH and AEH as a function of the normalized $S\to R$ link distances of $R_1$ and $R_2$. The  study is carried out for the NG and HI noise environments, considering Model I, under the assumption of both low (15~dB) and high (40~dB) SNR regimes. As it can be readily observed for all four relaying schemes, in general, the PEP increases as $d_{sr,1}$ and $d_{sr,2}$ increase, i.e., the distance between the source and the two relays increases. As explained earlier, this is because the farther away the two relay nodes are from the source node, the larger the experienced path-loss is,  leading to less signal power to be received at $R_n$. Accordingly, the received signal power at the destinations node is poor, yielding inferior PEP performance.  This result is in accordance with the majority of the research work in the literature of SWIPT relaying networks \cite{Nasir2013,Rabie1,Ojo,Rabie,Liu} and the references therein, where it is demonstrated that the best performance of the network was achieved when the relay nodes are located closer to the source node than the destination node. In our work, we demonstrate that this finding also holds when the network is operating under the impulsive noise. Moreover, we notice that in the case of low SNR regime (SNR=15dB), which is included in the flat region of the PEP performance under the HI noise, the PEP performance does not notably change with the change in the distance and that the performance is irrespective of the adopted relaying schemes. However, a rather more noticeable change is observed in the high SNR regime. This is in contrast to the NG noise environment case, where more rapid improvements can be seen at both low and high SNR regimes as the relays move closer to the source. Therefore, it turns out that moving the relays closer to the source is more rewarding in the NG noise environment. It can be further deduced from Fig. \ref{f6} that the performance gap between the four analyzed relaying schemes is more pronounced in the NG noise environment in the high SNR scenario. Finally, one can observe that the PEP performance does not notably change by increasing $d_{sr,1}$ and $d_{sr,2}$ beyond a certain value ($d_{sr,n}>0.8$), since as the relays get closer to the destination, smaller values of harvested energy are required to support the reliable communication through the $R_n\to D$ link. A similar conclusion can be drawn for all the presented relaying techniques for EH relays which are solely powered by the source. This suggests that the harvested energy at the relay nodes is the dominant performance limiting factor, rendering the $R_n\to D$ link to be the bottleneck of the system performance.   
\begin{figure}[!t]
\centering
   \includegraphics[width=3.7in]{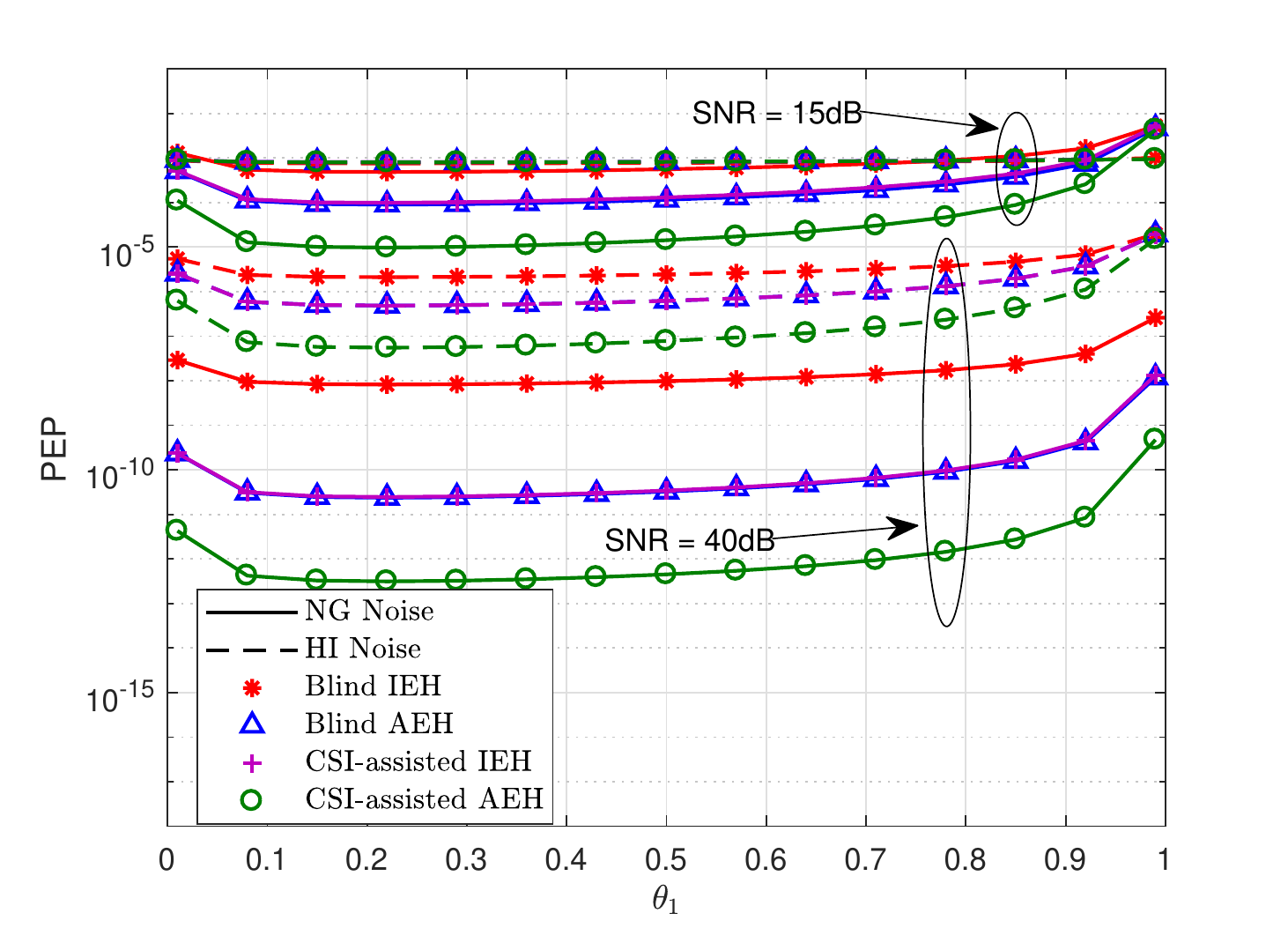}
  \caption{PEP performance with respect to the PS factor $\theta_1$ at relay $R_1$ over NG and HI noise under Model I, where $\theta_2=\theta_1$.}
   \label{f7a} 
 \vspace*{-0.3cm}
\end{figure}
\begin{figure}[!t]
\centering
   \includegraphics[width=3.7in]{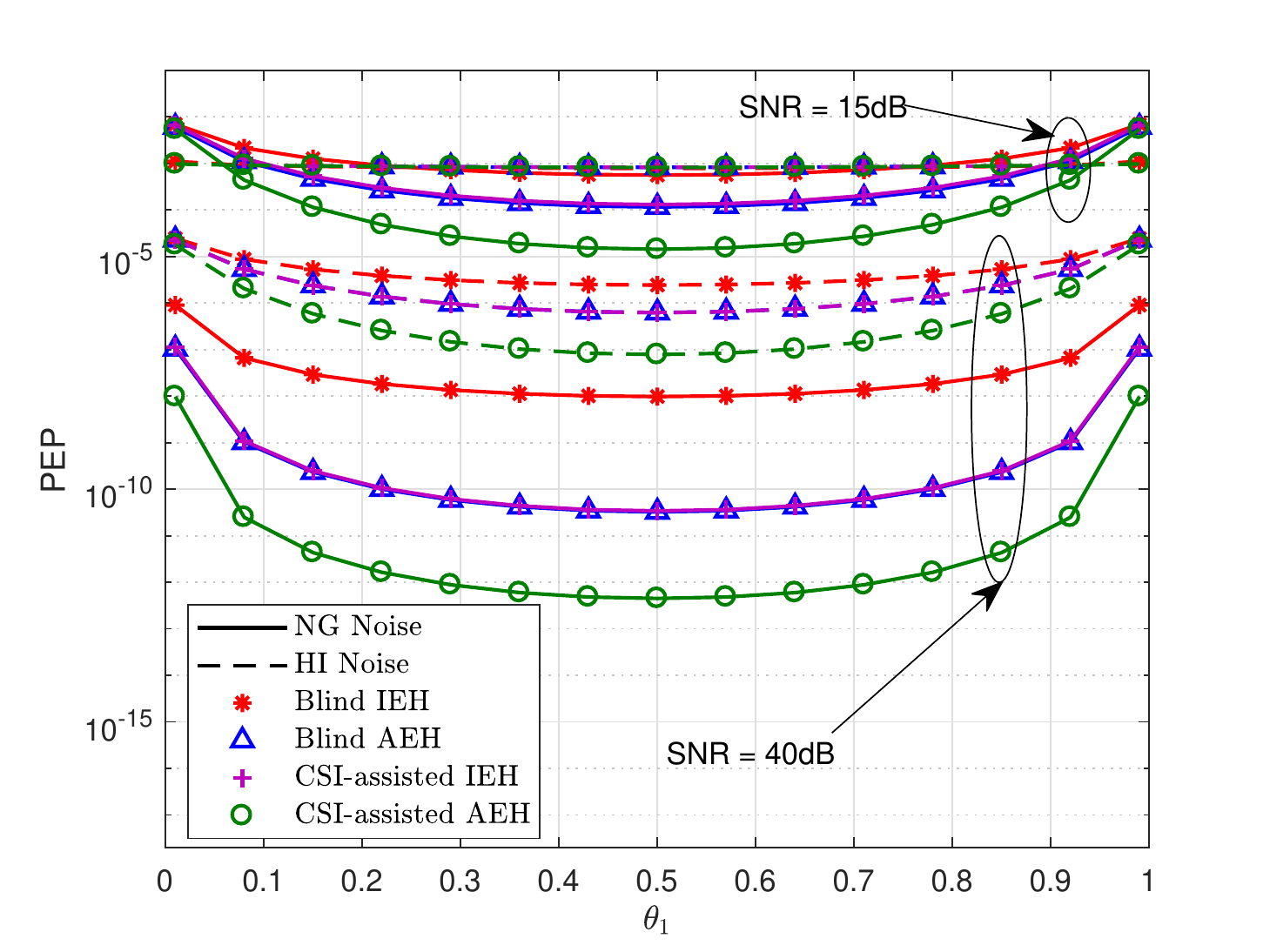}
  \caption{PEP performance with respect to the PS factor $\theta_1$ at relay $R_1$ over NG and HI noise under Model I, where $\theta_2=1-\theta_1$.}
   \label{f7b} 
 \vspace*{-0.6cm}
  \end{figure}
\par In Figs. \ref{f7a} and \ref{f7b}, we investigate the impact of the PS factor $\theta_n$ at the relays on the associated PEP performance of the competing relaying techniques for NG and HI noise environments under Model I. The examination is carried out for low and high SNR regimes.  Furthermore, in our work, we consider two scenarios for the PS factor of the two relays. The first scenario is depicted in Fig. \ref{f7a}, where we plot the PEP performance as a function of the PS factor of relay $R_1$, $\theta_1$, and we set the PS factor at the second relay $R_2$ to be $\theta_2=\theta_1$. In the second scenario, illustrated in Fig. \ref{f7b}, we set $\theta_2=1-\theta_1$. This is done to provide a deeper understanding on the behaviour of the system when equal or different power settings are imposed on the two relays. Interestingly, one can arrive at the same observation on the PEP performance from Fig. \ref{f7a} and  Fig. \ref{f7b}. Specifically, it is noted that the PEP performance is insensitive to the change in the value of the PS factors at the two relays in the HI noise environment under the low SNR assumption due to the detrimental effects of the impulsive noise. On the contrary,  it is demonstrated that for the other three scenarios (low SNR with NG noise and low and high SNR with NG and HI noise), there exists an optimal value for the PS factor that minimizes the PEP for the scenario in  Fig. \ref{f7a}. This stems from the fact that when the value of $\theta_n, n\in\lbrace{1,2\rbrace}$ is smaller than the optimal, there is less power available for EH. Consequently, less transmission power is available at the two relay nodes causing the performance to deteriorate gradually. On the other hand, as the value of $\theta_n$ increases beyond the optimal value, more power is spent on EH at the expense of the power available for data transmission which considerably degrades the PEP performance. This phenomenon is expected, since the performance of dual-hop systems is constrained by the quality of the weakest hop \cite{Muhaidat2}. Comparing the two setups, we observe from Fig. \ref{f7a} that the minimum PEP performance is attained when $\theta_1=\theta_2=0.22$. However, when the PS factors are different, we observe from Fig. \ref{f7b} that the minimum PEP is achieved for $\theta_1=\theta_2=0.5$. This finding suggests that allocating equal PS factors displays a performance gain gap over the non-equal PS factors at the two relays. A final observation for both Fig. \ref{f7a} and \ref{f7b} is that when blind IEH-relaying is adopted, varying $\theta_n$ only makes a rather small change to the PEP performance. This trend is similar for all the examined noise and SNR scenarios. The aforementioned two scenarios imply that the PS factor for EH must be optimized for best performance. 
 \begin{figure}[!t]
\centering
   \includegraphics[width=3.7in]{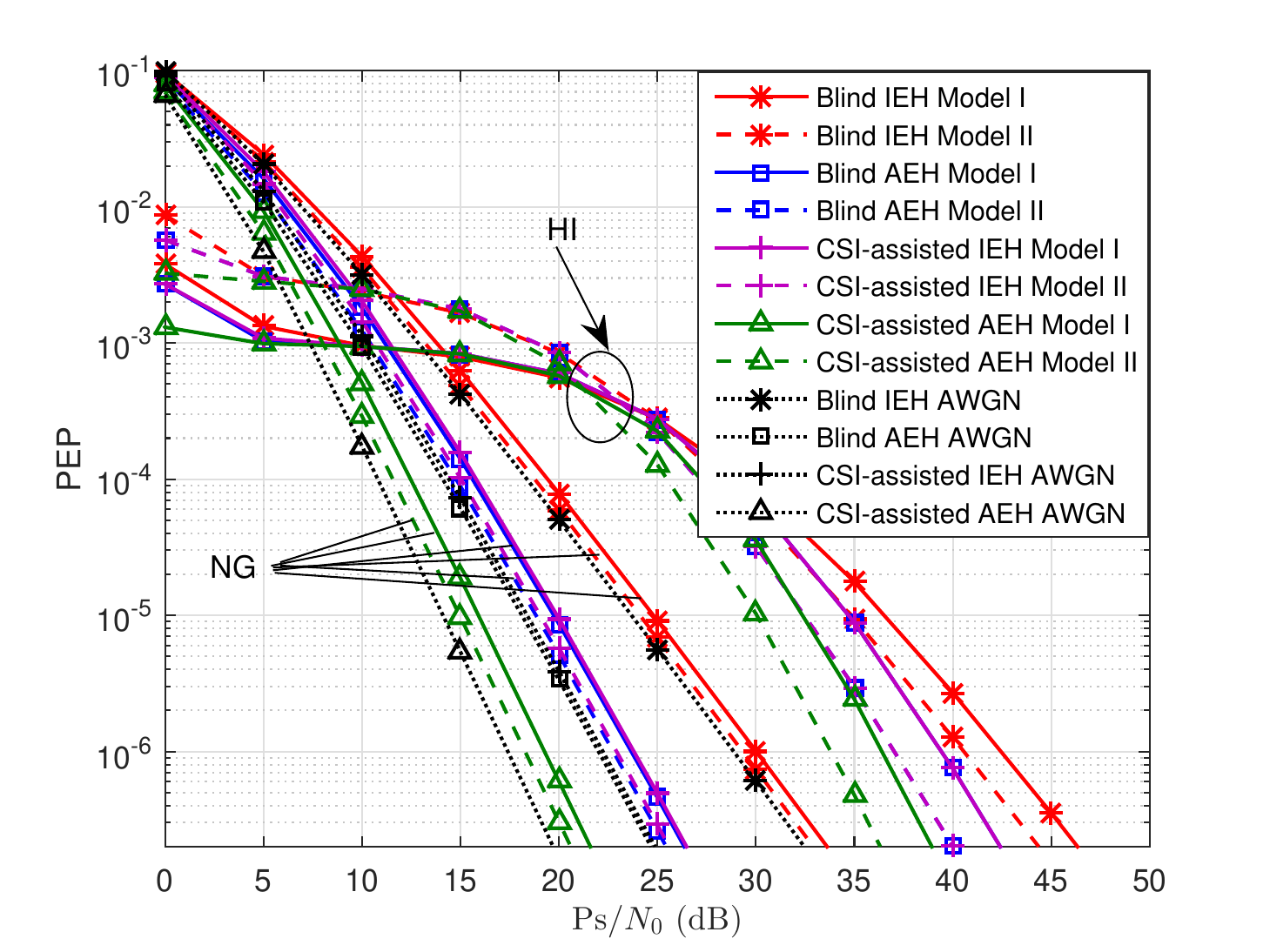}
  \caption{PEP performance over NG and HI noise under Model I and Model II.}
   \label{f8} 
\end{figure} 
\par To address the effect of the spatial independence, we plot in Fig. \ref{f8} the PEP performance for Models I and II under both NG and HI noise environments against the AWGN benchmark case. It is recalled that Model I refers to the case when the same set of interfering sources affects the relay and destination nodes together, while Model II refers to the case when different sets of interfering sources affect the relay and destination nodes.  Fig. \ref{f8} illustrates that when the noise is HI, Model I outperforms Model II in the sufficiently low SNR regime (SNR $<22$dB). This behaviour is reversed in the higher SNR region and the performance over Model II becomes superior to that exhibited by Model I. On the other hand, both models exhibit a similar performance in the NG noise over the whole inspected SNR region. These results are in accordance with the ones reported in \cite{Al-Dharrab}.
\section{Conclusions}\label{sec:conc}
In this paper, we have investigated the performance of distributed Alamouti codes for SWIPT AF relaying systems in the presence of MCA noise. Assuming the PS receiver architecture, we have derived novel closed-form PEP expressions which are then exploited to provide detailed performance comparisons among the four relaying techniques under consideration. Besides the fact that our results are accurate and mathematically tractable, they provide efficient means for the design and evaluation of SWIPT relaying networks in practical scenarios where impulsive noise is present. In particular, the proposed analytical model is exploited to study the diversity gains of blind AF and CSI-assisted AF schemes considering AEH and IEH. In addition, we have illustrated that the performance of CSI-assisted AEH-relaying is superior to that exhibited by the other three relaying techniques,  achieving the highest diversity order of 3. Furthermore, we have demonstrated that the performance loss incurred by the severity of noise impulsiveness increases with the diversity order and that the performance of the system in the low and medium SNR regions depends on the impulsive nature of the noise, resulting in different diversity orders to dominate the performance. Significant performance gains have been observed by locating the relays close to the source, offering a potential solution to mitigate the deleterious effect of MCA noise. Our results highlight the importance of accurately characterising the performance of the system for the successful implementation of SWIPT  relay networks in the presence of impulsive noise.
\vspace*{-0.5cm}
\appendices
\newcounter{tempequationcounter8}
\begin{figure*}[b]
\normalsize
\hrulefill
\setcounter{equation}{48} 
\begin{align}\label{I0}
I_0=-\frac{\xi_n}{B_n}\int_0^\infty\frac{\text{exp}\left(\frac{-t}{\xi_n}\right) dt}{\left(t+\left(\frac{\xi_n^2+\xi_n\sqrt{\xi_n^2-4B_n}}{2B_n}\right)\right)\left(t+\left(\frac{\xi_n^2-\xi_n\sqrt{\xi_n^2-4B_n}}{2B_n}\right)\right)} \nonumber \\
-\frac{\xi_n}{B_n}\int_0^\infty \frac{t\quad\text{exp}\left(\frac{-t}{\xi_n}\right) dt}{\left(t+\left(\frac{\xi_n^2+\xi_n\sqrt{\xi_n^2-4B_n}}{2B_n}\right)\right)\left(t+\left(\frac{\xi_n^2-\xi_n\sqrt{\xi_n^2-4B_n}}{2B_n}\right)\right)}
\end{align}
 \setcounter{equation}{\value{equation}}
\vspace*{4pt}
\vspace*{-0.3cm}
\end{figure*} 
\newcounter{tempequationcounter9}
\begin{figure*}[b]
\normalsize
\hrulefill
\setcounter{equation}{49} 
\begin{align}\label{pepprop3}
P(\bold{s}\to\hat{\bold{s}}||h_{sr_1}|^2,|h_{sr_2}|^2)\leq\sum_{m=0}^{M}\alpha_m\left(\frac{\Delta_s}{4\beta_m N_0}+1\right)^{-1}\quad \quad\quad \quad\quad \quad\quad \quad\quad \quad\quad \quad\quad \quad\nonumber \\
\times \left(\frac{\epsilon_1\zeta_1/(4\beta_m N_0))|h_{sr_1}|^2}{\xi_1|h_{sr_1}|^2+1}+1\right)^{-1}\left(\frac{(\epsilon_2\zeta_2/(4\beta_m N_0))|h_{sr_2}|^2}{\xi_1|h_{sr_2}|^2+1}+1\right)^{-1}.
\end{align} 
\setcounter{equation}{\value{equation}}
\vspace*{4pt}
\vspace*{-0.3cm}
\end{figure*} 
\setcounter{equation}{42} 
 \section{Proof of Proposition 1}
  \label{Appendix A}
Starting from the conditional PEP expression in \eqref{chernoff}, we take the expectation with respect to fading coefficients $|h_{sd}|^2$, $|h_{r_1d}|^2$, and $|h_{r_2d}|^2$, which follow an exponential distribution, resulting in
\begin{align}
P(\bold{S}\to\hat{\bold{S}}||h_{sr_1}|^4,|h_{sr_2}|^4)&\leq\sum_{m=0}^{M}\alpha_m\left(\frac{\Delta_s}{4\beta_m N_0}+1\right)^{-1}\nonumber \\
&\times\prod_{n=1}^2 \left(\frac{\epsilon_n\Phi_n^2 }{4\beta_m N_0}|h_{sr_n}|^4+1\right)^{-1}.
\end{align}
Performing an expectation with respect to the random variables $|h_{sr_1}|^4$, $|h_{sr_2}|^4$, which also follow an exponential distribution, yields the unconditional PEP, which is written as\\
\begin{align}\label{uncondPEP1}
P(\bold{s}\to\hat{\bold{s}})&\leq\sum_{m=0}^{M-1}\alpha_m\left(\frac{\Delta P_s}{4\beta_m N_0}+1\right)^{-1} \nonumber \\
&\times\prod_{n=1}^2\int_0^\infty\left(\frac{\epsilon_n \Phi_n^2 }{4\beta_m N_0}t^2+1\right)^{-1}\text{exp}(-t)  dt,
\end{align}where $t$ is the integration variable and $n\in\lbrace{1,2\rbrace}$. Using the equality in \cite[Eq. (8.4.2.5)]{Prudnikov} to express the first integrand of \eqref{uncondPEP1} as \\
\begin{equation}
\left(\frac{\epsilon_n \Phi_n^2}{4\beta_m N_0}t^2+1\right)^{-1}=G_{1, 1}^{1, 1}\left[\frac{\epsilon_n \Phi_n^2}{4\beta_m N_0}t^2 \Big \vert \ {0 \atop 0} \right], \quad n\in\lbrace{1,2\rbrace}, 
\end{equation}\\then making use of the equality $e^{-t}=G_{0, 1}^{1, 0}\left[t\ \vert \ {- \atop 0} \right]$\cite[Eq. (8.4.3.1)]{Prudnikov} to rewrite the second integrand in \eqref{uncondPEP1}, the unconditional PEP can be derived in a closed-form as in \eqref{finalPEPmodIAPS} by exploiting the integral identity \cite[Eq. (2.24.1.2)]{Prudnikov}.\\
\section{Proof of Proposition 2}
  \label{Appendix B}
In order to derive the PEP expression, we first obtain the exact PDF of the RV $Z_n$. It is recalled that RVs $X$ and $Y$\footnote{We drop in the proof the index $n$ for the convenience of analysis.} are independent RVs drawn from the exponential distribution. Therefore, their joint PDF is $f_{X,Y}=e^{-x-y}$ \cite{papoulis}. Expressing $X$ and $Y$ in terms of $U$ and $V$ as $X=(V-1)/\xi$ and $Y=U\xi^2/(V-1)^2$, then with the help of the Jacobian transformation method \cite{papoulis}, $(X,Y)$ are transformed to $(U,V)$. Consequently, the PDF of $(U,V)$ is obtained as \\
\begin{equation}\label{pdfuv}
f_{U,V}=J_d f_{X,Y}\left(\frac{(v-1)}{\xi},\frac{ub^2}{(v-1)^2} \right), 
\end{equation}where $J_d=-\xi/(V-1)^2$ is the Jacobian of the transformation. Then using \cite[Eq. (6.60)]{papoulis} and \eqref{pdfuv}, and after some algebraic manipulations, the exact PDF of $Z$ is derived as\\ \\
\begin{align}\label{pdfz}
f(z)=\int_1^\infty v f_{U,V}(vz,v) dv \quad\quad\quad\quad\quad\quad\quad\quad \quad \quad  \quad \quad  \nonumber \\
=-\int_0^\infty \frac{\xi(t+1)}{t^2} f_{X,Y}\left(\frac{t}{\xi},\frac{(t+1)z\xi^2}{t^2}\right) dt\quad \quad \quad  \quad \nonumber \\
 =-\int_0^\infty \frac{\xi (t+1)}{t^{2}}\text{exp}\left(-\frac{t}{\xi}-\frac{\xi^2  (t+1) z}{t^2}\right)dt, \quad \quad\quad
\end{align}\\where the second equality in \eqref{pdfz} stems from the fact that $v>1$, as shown in \eqref{pdfuv}. To the best of the authors' knowledge, the integral in \eqref{pdfz} does not lend itself to a closed-form. However, we can obtain the exact PEP expression in a closed-form by substituting \eqref{distIPS} in the conditional PEP expression given in \eqref{chernoff}. Then, the desired unconditional PEP expression is deduced in \eqref{pepIEHIPS} by taking the expectation with respect to the RVs $|h_{sd}|^2, Z_1$ and $Z_2$, where we used the fact that the PDF of $|h_{sd}|^2$ follows the exponential distribution and that the PDF of each of $Z_1$ and $Z_2$ is computed using \eqref{pdfz}, yielding
\begin{align}\label{peppdfz}
P(\bold{s}\to\hat{\bold{s}})\leq\sum_{m=0}^{M-1}\alpha_m\left(\frac{\Delta P_s}{4\beta_m N_0}+1\right)^{-1}\quad\quad\quad\quad\quad\quad\quad\quad\nonumber\\
\times\prod_{n=1}^2\underbrace{\int_0^\infty -\xi_n (t+1) t^{-2}\left(\frac{\xi_n^2 (t+1)}{t^2}+B_n\right)^{-1}\text{exp}\left(\frac{-t}{\xi_n}\right) dt}_{I_0},
\end{align}where $B_n=(\epsilon_n\zeta_n)/(4\beta_m N_0)$. Finally, by rewriting $I_0$ as \eqref{I0} at the bottom of this page, followed by some algebraic manipulations, and invoking \cite[Eq. (3.354.3)]{Rizhik} and \cite[Eq. (3.354.4)]{Rizhik}, the integral in \eqref{peppdfz} is obtained in a closed-form as in \eqref{pepIEHIPS}.
\vspace*{-0.3cm}
\section{Proof of Proposition 3}
  \label{AppendixC}
Substituting \eqref{distIPS2} in the conditional PEP expression in \eqref{chernoff}, then taking the expectation with respect to fading coefficients $|h_{sd}|^2$, $|h_{r_1d}|^2$, and $|h_{r_2d}|^2$, which follow an exponential distribution, to yield \eqref{pepprop3} at the bottom of this page. Performing an expectation with respect to the random variables $|h_{sr_1}|^2$, $|h_{sr_2}|^2$, which also follow an exponential distribution, yields the unconditional PEP which is written as
\setcounter{equation}{50} 
\begin{align}\label{uncondPEP}
P(\bold{s}\to\hat{\bold{s}})\leq\sum_{m=0}^{M-1}\alpha_m\left(\frac{\Delta P_s}{4\beta_m N_0}+1\right)^{-1}\quad \quad\quad\quad \quad\quad\quad \quad\quad\nonumber \\
\times\prod_{n=1}^2\underbrace{\int_0^\infty\left(\frac{(\epsilon_1\zeta_1/(4\beta_m N_0))|h_{sr_1}|^2}{\xi_1|h_{sr_1}|^2+1}+1\right)^{-1}\text{exp}(-t) dt}_{\Upsilon},
\end{align}where $t$ is the integration variable and $n\in\lbrace{1,2\rbrace}$. To solve the integral $\Upsilon$, we perform simple algebraic manipulations to get
\begin{align}
\Upsilon&=\int_0^\infty\left(\xi_n t+1\right)\left(\left(\frac{\epsilon_n\zeta_n}{4\beta_m N_0}+\xi_n\right) t+1\right)^{-1}\text{exp}(-t) dt \nonumber \\
&=\underbrace{\int_0^\infty \left(\left(\frac{\epsilon_n\zeta_n}{4\beta_m N_0}+\xi_n\right) t+1\right)^{-1}\text{exp}(-t) dt}_{I_1}\nonumber \\
&+\xi_n\underbrace{\int_0^\infty t\left(\left(\frac{\epsilon_n\zeta_n}{4\beta_m N_0}+\xi_n\right) t+1\right)^{-1}\text{exp}(-t) dt}_{I_2} \label{47a}. 
\end{align}
Then, with the aid of the equality in \cite[Eq. (8.4.2.5)]{Prudnikov}, followed by applying the transformation \cite[Eq. (8.2.2.14)]{Prudnikov}, the first and second integrands of $I_1$ and $I_2$, respectively, are expressed in terms of their Meijer G-function representations as\\
\begin{equation}
\left(\gamma_n t+1\right)^{-1}=G_{1, 1}^{1, 1}\left[\frac{1}{\gamma_n t } \ \Big \vert \  {1 \atop 1} \right].
\end{equation}where $\gamma_n=\left(\frac{\epsilon_n\zeta_n}{4\beta_m N_0}+\xi_n\right)$. Similarly, the second and third integrands of $I_1$ and $I_2$, respectively, are rewritten by making use of the equality $e^{-t}=G_{0, 1}^{1, 0}\left[t \  \vert \  {- \atop 0} \right]$\cite[Eq. (8.4.3.1)]{Prudnikov}, yielding
\begin{align}
\Upsilon&=\int_0^\infty G^{1,1}_{1,1}\left[\gamma_n t \ \Big\vert \ {0 \atop 0}\right]G^{1,0}_{0,1}\left[t \ \Big\vert \ {- \atop 0}\right]dt \nonumber \\
&+\xi_n\int_0^\infty t G^{1,1}_{1,1}\left[\gamma_n t \ \Big\vert \ {0 \atop 0}\right]G^{1,0}_{0,1}\left[t \ \Big\vert \ {- \atop 0}\right]dt. \label{b47}
\end{align}
Then, by exploiting the integral identity \cite[Eq. (3.356.4)]{Prudnikov}, followed by performing some algebraic manipulations, $\Upsilon$ can be derived in a closed-form as 
\begin{align}
\Upsilon&=\gamma_n^{-1}G^{1,2}_{2,1}\left[\gamma_n \ \Big\vert \ {1, 1 \atop 1}\right] +\xi_n \gamma_n^{-2}G^{1,2}_{2,1}\left[\gamma_n \ \Big\vert \ {1, 2 \atop 2}\right] \label{c47}
\end{align}
Finally, after substituting \eqref{c47} in \eqref{uncondPEP}, the desired result in \eqref{PEPAEHIPS} is derived.

\balance

\bibliographystyle{IEEEtran}
\bstctlcite{BSTcontrol}
\bibliography{Reflist}
\end{document}